# Title:

# Tailoring wetting properties of organic hole-transport interlayers for slot-die coated perovskite solar modules

Authors:

T.S. Le [a],[1] I.A. Chuyko[a,b],[1] L.O. Luchnikov[a], K.A. Ilicheva[a], P.O. Sukhorukova [a,b], D.O. Balakirev[b], N.S. Saratovsky [b], A.O. Alekseev[a], S.S. Kozlov[c], D.S. Muratov[d], V.V. Voronov[a], P.A. Gostishchev[a], D. A. Kiselev[e], T.S. Ilina[e], A.A. Vasilev[f], A. Y. Polyakov[f], E.A. Svidchenko,[b] O.A. Maloshitskaya[g], Yu. N. Luponosov [b]* and D.S. Saranin [a]*

[a]LASE – Laboratory of Advanced Solar Energy, National University of Science and Technology "MISiS", Leninsky Prospect 4, 119049, Moscow, Russia

[b]Enikolopov Institute of Synthetic Polymeric Materials of the Russian Academy of Sciences (ISPM RAS), Profsoyuznaya St. 70, Moscow, 117393, Russia

[C]Laboratory of Solar Photoconverters, Emanuel Institute of Biochemical Physics, Russian Academy of Sciences, 119334 Moscow, Russia

[d]Department of Chemistry, University of Turin, 10125, Turin, Italy

[e]Laboratory of Physics of Oxide Ferroelectrics, Department of Materials Science of Semiconductors and Dielectrics, National University of Science and Technology MISIS, Moscow 119049, Russia

[f]Department of Semiconductor Electronics and Semiconductor Physics, National University of Science & Technology MISIS, 4 Leninsky Ave., Moscow, 119049, Russia

[g]Moscow State University, Chemistry Department, 1/3 Leninskie Gory, Moscow, 119991, Russia

[1]The authors contributed equally to this work.

Corresponding author E-mail addresses: luponosov@ispm.ru (Yu.N. Luponosov), saranin.ds@misis.ru (D.S. Saranin).

## Abstract:

Among the various technologies for scaling up halide perovskite solar modules (PSMs), slot-die coating is one of the most promising. The use of self-assembled monolayers (SAMs) with anchoring groups was considered as an effective approach for interface engineering in perovskite solar cells with metal oxide charge transporting layers. However, the coating of SAM layers in PSMs by means of a slot-die is a challenging process due to the low viscosity of the solutions and the low wettability of the films. In this study, we integrate a triphenylamine-based polymer, pTPA-TDP, blended with SAM based on 5-[4-[4-(diphenylamino) phenyl] thiophene-2-carboxylic acid (TPATC), to address the challenges of uniform slot-die coating and interface passivation in large-area modules. We fabricated p-i-n oriented PSMs on 50x50 mm$^2$ substrates (12-sub-cells) with NiO hole transport layer (HTL) and organic interlayers for surface modification. Wetting angle mapping demonstrated that ununiform regions of the slot-die coated SAM have hydrophobicity with contact angle values up to 90°, causing fluctuations in absorber thickness and the presence of macro-defects at buried interfaces. The incorporation of the blended interlayer to NiO/perovskite junction homogenized the surface wettability (contact angle=40°) and mitigated lattice strain in the absorber. This enabled the effective use of SAM properties on a large-area surface, improving energy level alignment and enhancing the power conversion efficiency (PCE) of the modules from 13.98% to 15.83% and stability (ISOS-L-2, $T_{80}$ period) from 500-1000 hours to 1630 hours. Investigation of PSMs upon cooling till -5 °C showed that the PCE increased by +0.19%/°C for samples with NiO HTL, while using SAM and blended interlayers raised the coefficient to ~0.40%/°C due to changes in activation energy and trap contributions to device performance across a wide temperature range. The results provide important insights into using SAM in slot-die coating technology for perovskite photovoltaics and highlight the complex effects on the operation of large-scale solar modules.

**Introduction:**

Halide perovskites (**HPs**) stand out as one of the most promising semiconductors for third-generation solar cells[1]. The $ABX_3$ structure in HP crystals is characterized by A-site cations $CH_3NH_3$ (MA), $CH(NH_2)_2$ (FA) or inorganic Cs; B-site cations, typically Pb or Sn; and X-site anions – I, Br, Cl. Strong optical absorption of HPs in the visible spectra[2] coupled with a tunability of band-gap energies and suppressed dynamics of non-radiative recombination processes[3] enables to achieve high power conversion efficiencies (**PCE**) of solar cells. As reported, lab-scale HP-based photovoltaic devices, with an active area of <1 cm², have reached record efficiencies of 26.1% under standard conditions[4]. Such performance is on par with commercial technologies like monocrystalline silicon and copper indium gallium selenide (CIGS)[4]. HP's technology leverages solution-processing, employing industrial techniques such as inkjet printing[5], slot-die[6] and spray coating[7]. This approach gives opportunities for cost-effective industrial production with reduced capital expenses (**CAPEX**)[8]. Perovskite solar cells (**PCSs**) consist of multilayer thin films, featuring a sub-micron thick absorber layer sandwiched between p-type and n-type transport layers. The architecture of PCSs has a p-i-n or n-i-p orientation, which is determined by the positioning of the charge-transporting layer (**CTL**) on the transparent conducting electrode (**TCE**), which facilitates the passing of light into the solar cell. Typically, the CTL applied to the surface of the TCE is based on a wide band-gap oxide fabricated via high-temperature decomposition of precursors[9] or deposition from nanoparticle dispersions[10,11]. $SnO_2$ and $TiO_2$ (n-type), as well as NiO[12] and $Nb_2O_5$ (p-type) provide high optical transparency and relevant energy level alignment with perovskite absorber for efficient charge collection. The chemical stability of the perovskite interface with CTLs is crucial for the long-term operation of solar cells[13]. HP absorbers can undergo decomposition when exposed to heating and illumination[14,15], leading to the formation of HI, $I_2$, volatile compounds, metallic lead, and other byproducts[16]. Defects in non-stoichiometric oxide CTLs could contribute to chemical activity at the interfaces. As reported[17,18], uncompensated nickel atoms in NiO can form chemical bonds with the organic cation in the perovskite molecule. This interaction triggers the decomposition of the absorber and corrosion in the device structure.

The use of self-assembled monolayer materials (SAM) based on organic conjugated molecules has been proposed as an effective strategy to passivate oxide CTL surfaces in PSCs [19,20]. SAMs are thin organized structures formed on the surface of inorganic materials by strong chemical interactions between functional groups of SAMs and oxygen or metal atoms of the surface. In general, SAM molecules consist of a terminal group responsible for interacting with the atmosphere or overlayer, an anchoring group responsible for forming chemical bonds, and a linker between these groups[21]. The formation of SAMs in PSCs serves to increase the stability and efficiency of the device by creating a mechanical barrier between the perovskite and the surrounding atmosphere and reducing the number of defects at the perovskite-substrate interface[22]. One of the possible mechanisms for passivation of the perovskite surface by SAM materials is the binding of insufficiently coordinated surface Pb atoms, which can accumulate free iodine formed during perovskite degradation, leading to an imbalance of iodine in the perovskite volume and the appearance of additional defects. Surface passivation is therefore a key factor in preventing the release of iodine from the perovskite and, consequently, the degradation of the device[23].

Up-scaling perovskite solar modules (**PSMs**) with SAM interlayers presents a complex technological challenge. To ensure effective functionality of SAMs, it is critical to provide uniformity of the ultrathin interlayer over large device areas. The recent reports presented the uniform deposition of SAM via thermal evaporation[24]. However, cost-effective and high-throughput production cycle require implementation of solution-processing techniques.

Employing conventional SAMs configurations (2PACz etc.) increases contact angle of surface wetting for solution processing of perovskite precursors. Liquid-phase crystallization of perovskite absorbers on the hydrophobic surfaces improves the average grain size and reduces the density of the structural defects. These improvements originated from the reduced concentration of nuclei in the wet film of perovskite solution during crystallization process on the surface modified with SAM. 2PACz ([2-(9H-carbazol-9-yl)ethyl]phosphonic acid) and MeO-2PACz ([2-(3,6-dimethoxy-9H-carbazol-9-yl)ethyl]phosphonic acid), have been widely adopted for tuning of energy levels in single junction and tandem devices[25,26]. However, the poor coverage of SAM interlayers from the perovskite precursor solution limits the up-scaling of PV devices. Addressing the wetting issue could advance application of SAM for large area solar modules. *Al-Ashoury and co-workers*[27] modified Me-4PACz with 1,6-hexylenediphosphonic acid (6dPA) to reduce the wetting angle of the surface with deposited SAM. The authors demonstrated a significant decrease in quantity of shorted pixels on 25x25 mm$^2$ substrate from 67% for unmodified surface to 8% for the sample with Me-4PACz:6dPA interlayer. *Zhang et al.*[28] introduced an effective method using an amphiphilic molecular hole transporter, (2-(4-(bis(4-methoxyphenyl)amino)phenyl)-1-cyanovinyl)phosphonic acid that forms superwetting underlayer for perovskite deposition. The improvement in surface wetting of hole transporting layers suppressed formation the defects at the buried interface and allowed to achieve the PCE of 25.15% for small-area devices. We found that the current literature lacks a comprehensive analysis of the application of SAM in industrial technological processes. Among the limited reports on scaling solar modules using SAMs[29–31], slot-die and doctor blade methods have been employed only for the perovskite absorbers, while the passivation interlayer was deposited via spin-coating or rinsing. Tailoring the wetting properties of SAMs deposited using scalable methods remains a crucial challenge that requires further investigation.

In this paper, we present a complex analysis of SAM application in the slot-die coating process of inverted perovskite solar modules with NiO HTL. Our results demonstrate that the high inhomogeneity typically associated with slot-die coated SAM can be mitigated through the use of a donor-acceptor triphenylamine-based polymer (pTPA-TDP). The incorporation of a mixed polymer-SAM interlayer has been shown to enhance the quality of the perovskite absorber without pinholes in perovskite modules (12 sub-cells) on 50 × 50 mm² substrates.

**Results and discussion**

In this study, we examined the impact of surface properties at the hole-transport interface on PSMs, which were fabricated using the slot-die coating method. We investigated p-i-n oriented devices with the following architecture (**fig.1(a)**): ITO(anode)/ NiO (hole-transport layer)/Organic interlayer/$Cs_{0.2}FA_{0.8}PbI_{2.9}Cl_{0.1}$(perovskite absorber)/$C_{60}$(electron transport layer)/Bathocuproine:$Ti_3C_2$(hole blocking interlayer/Bismuth-Copper (cathode). The electronic supplementary material (**ESI**) provides a detailed description of the experimental procedures. The hole-transport layer, organic interlayer, and perovskite absorber were sequentially deposited using the slot-die method on ITO substrates. All slot-die coating steps involved in fabricating the PSMs were performed at room temperature, within a range of 20–30°C, and with a relative humidity (**RH**) of 10–40%, under ambient conditions. Briefly, NiO HTL was slot-die coated following the decomposition and oxidation of the solution-processed $NiCl_2$, as shown in our previous work[32]. The $NiCl_2$ precursor was deposited at a pre-heating temperature of 80°C, employing the following parameters: a printing speed of 15 mm/s; syringe rate of 8 μL/s and meniscus gap -150 μm. After placing all the as-coated films on a hot plate set at 120°C, the temperature was incrementally raised to 300°C to facilitate further annealing for 1 hour(h). Organic interlayers were deposited on the NiO surface to modify the HTL and perovskite absorber interface. We focused on studying the SAM (TPATC) and the blended interlayer of SAM with polymer - pTPA-TDP (**Fig. 1**). The synthetic procedure, thermogravimetric properties and optical absorption characterization of the materials are described in the Electronic supplementary information (**ESI**) (**Fig. S1-S6**). In essence, the properties of a mixture are typically determined by an additive function of the properties of the individual two materials.

Our recent research[33] indicates that TPATC-SAM is a promising passivation of NiO interface in p-i-n oriented PCS due to complementary optoelectronic and transport properties. Spin-coating fabrication of perovskite on the SAM surface has demonstrated high efficiency for lab-scale samples and modules. In contrast, the slot-die process has radically different wet film formation kinetics, operating time windows, and crystallization specifics during post-processing. In layer-by-layer deposition for p-i-n PSMs, the surface properties of HTLs significantly impact the morphology and quality of the perovskite absorbers. Fabrication of a uniform perovskite film necessitates high homogeneity in wettability across the substrate of the solar module. This uniformity is crucial both during the wet film pulling from the meniscus and throughout the subsequent crystallization process from the solution. During solution deposition, TPATC can form aggregates due to its low viscosity and high intermolecular interactions, resulting in SAM molecules not being covalently bonded to NiO over the entire surface. In these areas, the surface properties of the films, such as hydrophobicity and surface energy, can change significantly. Increasing the solution's viscosity and reducing the local TPATC concentration are keys to solving this problem. The polymer solution differs from the SAM solution primarily due to its higher viscosity. In order to improve the homogeneity of organic interlayers with SAM properties, we have developed a blend based on TRATC as SAM and a hole transporting polymer pTPA-TDP, which are largely similar in chemical structure and composition[34]. The pTPA-TDP polymer exhibits high thermal stability (decomposition temperature exceeds 500 °C), good solubility, deep HOMO energy levels (-5.35 eV), and relevant hole mobility (up to $7 \times 10^{-5}$ cm$^2$ V$^{-1}$ s$^{-1}$) without any post-treatment in thin-films [34]. The bare solution of SAM and its mixture with pTPA-TDP (1:1 by mass ratio) were slot-die coated at a speed of 15 mm/s and a syringe rate of 10 µL/s at room temperature. The as-coated substrates were placed in a chamber for vacuum-assisted drying for 5 seconds. The perovskite precursor solution was coated at a speed of 15 mm/s and a solution feed rate of 12 µL/s. Then, we transferred the wet films to a vacuum chamber and kept them under high vacuum for 2 minutes. Subsequently, the films were annealed in air at 105°C for 30 minutes.

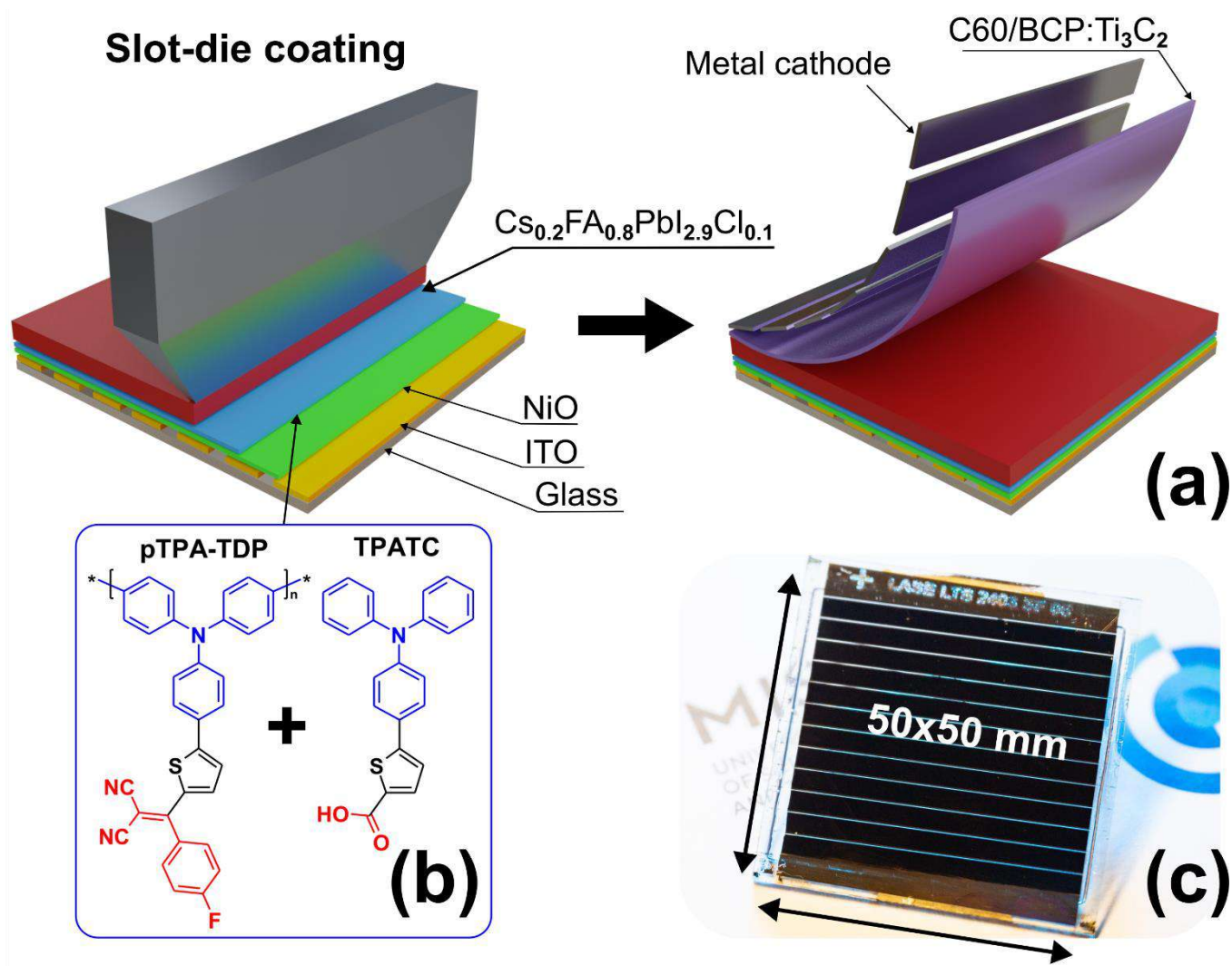

**Figure 1** – Sketch schematics of the slot die coated perovskite solar modules (a) the chemical structures of the SAM (TPATC) and polymer (pTPA-TPD) used for medication of hole transport interface (b) the photo-image of the fabricated perovskite solar module (c)

We employed stylus profilometry to estimate the thickness of the fabricated organic interlayers. Following statistical analysis of data collected from 10 points on a substrate (ITO-glass with NiO HTL), we found the average thickness of the slot die-coated TPATC was ~8 nm (±1.2 nm), while the Blend configuration averaged ~10 nm (±1.5 nm). We also analyzed the optical properties of the obtained samples. **Fig. S7** (**ESI**) presents the light transmission spectra for thin-films of the organic interlayers. Estimating absorption is essential for interlayers at the HTL interface in p-i-n devices, due to the potential for parasitic absorption effects, as light passes through the multilayer device stack before reaching the perovskite film. TPATC and blended samples demonstrated maximum losses at around 415 nm in the short-wavelength region, which is consistent with absorbance data (**fig.S5**) For the TPATC interlayer, transmittance losses varied between 2% and 4% in the 450–800 nm range, whereas the blend configuration exhibited increased losses of 3%–5% within the 500–600 nm wavelength range.

To analyze the surface properties of NiO with organic interlayers, we assessed the uniformity of the contact wetting angle (**CWA**) across 50x50 mm² substrates, as shown in **Fig.2** and **Fig.S8** (**ESI**). We applied deionized water droplets (5 -10 µL) onto the substrates, maintaining a spacing of 6.25 mm. Analysis of the wetting angle distribution maps for the NiO configuration (**fig.S2(a)**) reveals low values, averaging 27º. The slot-die coating of NiO thin films provided high surface wetting uniformity, as evidenced by a standard deviation of 3.2º. In the NiO/TPATC configuration, we observed a significant increase in the CWA accompanied by notable inhomogeneities. The CWA values ranged from 1º to an exceptionally high 90º, with the standard deviation increasing to 19.3º. It is evident that the heterogeneity in the wetting of the TPATC interlayer affects the fabrication of high-quality, pin-hole free perovskite absorber films. We attribute the local variations in CWA of the TPATC surface to capillary effects arising during the formation of a wet film in the slot-die printing process. The use of blend mitigated the formation of local inhomogeneities associated with SAM during slot-die coating **fig.2(b)**. Compared to bare NiO HTL, the blended organic interlayer demonstrated a relevant increase of CWA to 40.7º, along with a reduced standard deviation of 3.5º. Hence, the optimal technique involves the blending with TPATC, which improves hydrophobicity and uniformity. This provides favorable surface properties for the deposition and crystallization of the perovskite absorber. To simplify the titles of the different HTLs configurations, we will designate the sample without organic interlayers as "NiO". The sample with an organic TPATC interlayer (NiO/TPATC) will be designated as "SAM", and the sample of NiO with a blend of TPATC and pTPA-TDP will be referred to as "Blend".

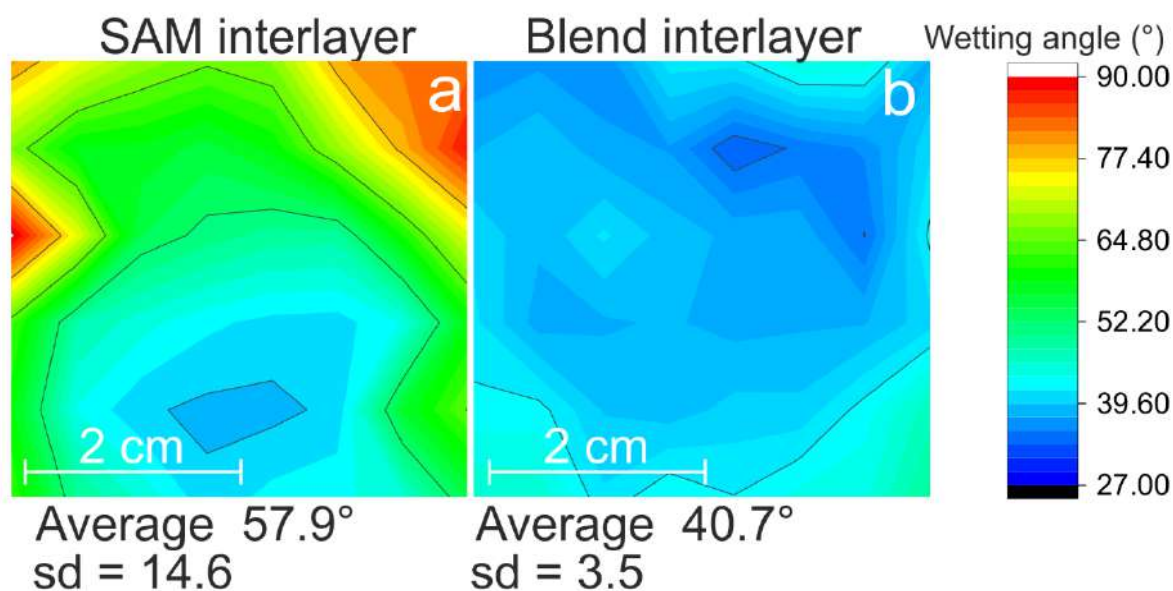

Figure 2 – The heat-map of CWA distribution across the 50 x 50 mm² ITO/NiO substrate with TPATC interlayer (a), and blended TPATC-pTPA-TDP (b)

Ununiform regions with significantly different wettability can strongly influence the optoelectronic properties of the HTL, as well as the structure and morphology of the perovskite absorber crystallized on the surface. Therefore, we investigated the changes in properties in different regions of the substrates, such as the center and edge regions, with respect to the schematics in **Fig. S9 (ESI)**. The work function ($W_f$) measurements performed using the KPFM method are shown in **Fig. 3**. NiO samples showed a uniform distribution of work function on the surface, with an average value of $W_f$ =4.977eV (**Fig. 3(a,e)**). For the SAM samples, huge variations were measured in the central and top-right areas (**Fig. 3(b,c,f,g)**). The inhomogeneities observed in the top-right region of SAM have an amplitude of up to 70 nm and can be attributed to the formation of TPATC clusters that have not interacted with the substrate via anchoring groups. These regions correspond to the maximum surface hydrophobicity (CWA~90°) and an increased $W_f$ = 4.919 eV. In contrast, the central area of the SAM sample shows uniform morphology and negligible changes in $W_f$ compared to NiO. For the Blend interlayer (**Fig. 3(d,h)**), $W_f$ increased to 5.036eV without extensive areas of inhomogeneity across the substrate surface. The histogram of the calculated $W_f$ data presented in **fig.S10** (**ESI**).

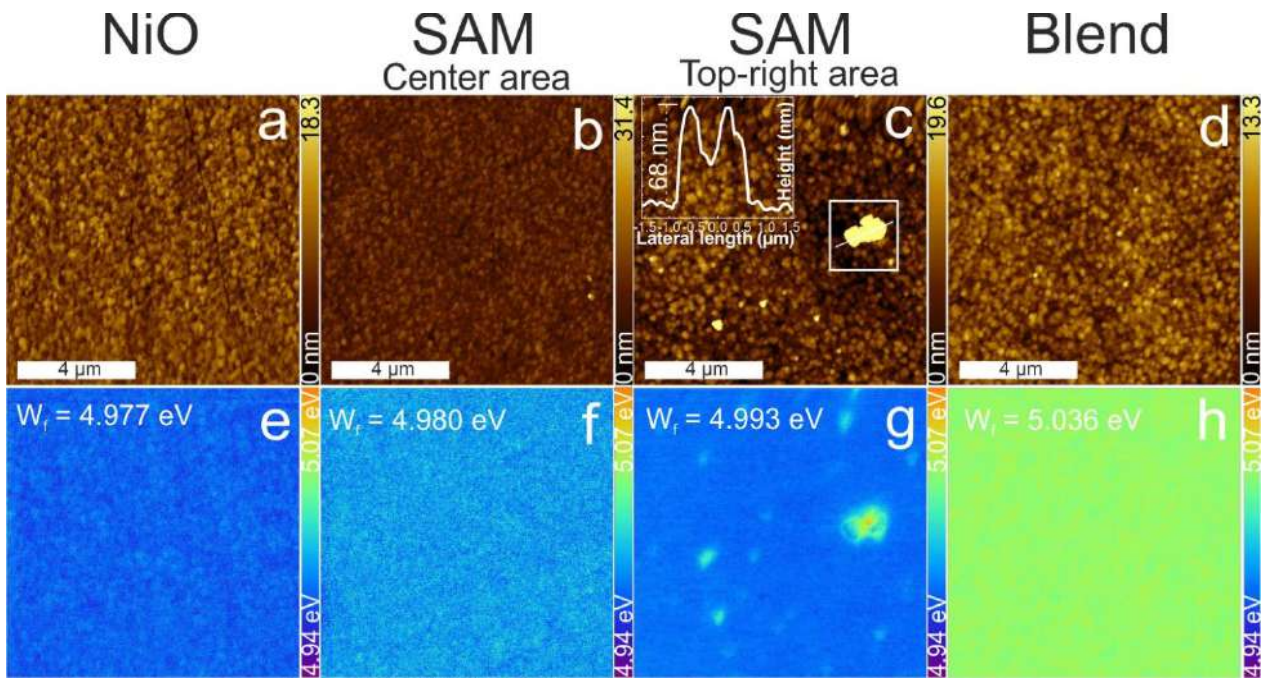

Figure 3 – morphology of NiO (a), SAM in central area (b), SAM in top-right area (c), and blend (d) thin films on ITO substrate and work function of corresponding layers (e-h)

The study of perovskite film surface morphology using an atomic force microscope is shown in **Fig. 3 (a-d)** and **fig.S11** (**ESI**). The average grain size on the surface of all samples was in the range of 160 ± 10 nm, with no significant differences across different configurations. However, macro-defects were observed when we investigated the buried interface of the absorber with HTLs (**Fig. 4**). The perovskite film crystallized on NiO, with an average CWA=26.5°, is characterized by the presence of voids and inclusions across all areas of the substrate. The central region of the SAM sample is defect free, while regions with a low wetting angle exhibit voids up to 112 nm in depth. Also, we found that grains with a diameter of up to 300 nm protrude at a height of 50 nm in the top right corner. Using Blend allows for the formation of a continuous, uniform film across the entire surface of the 50 x 50 mm² substrate.

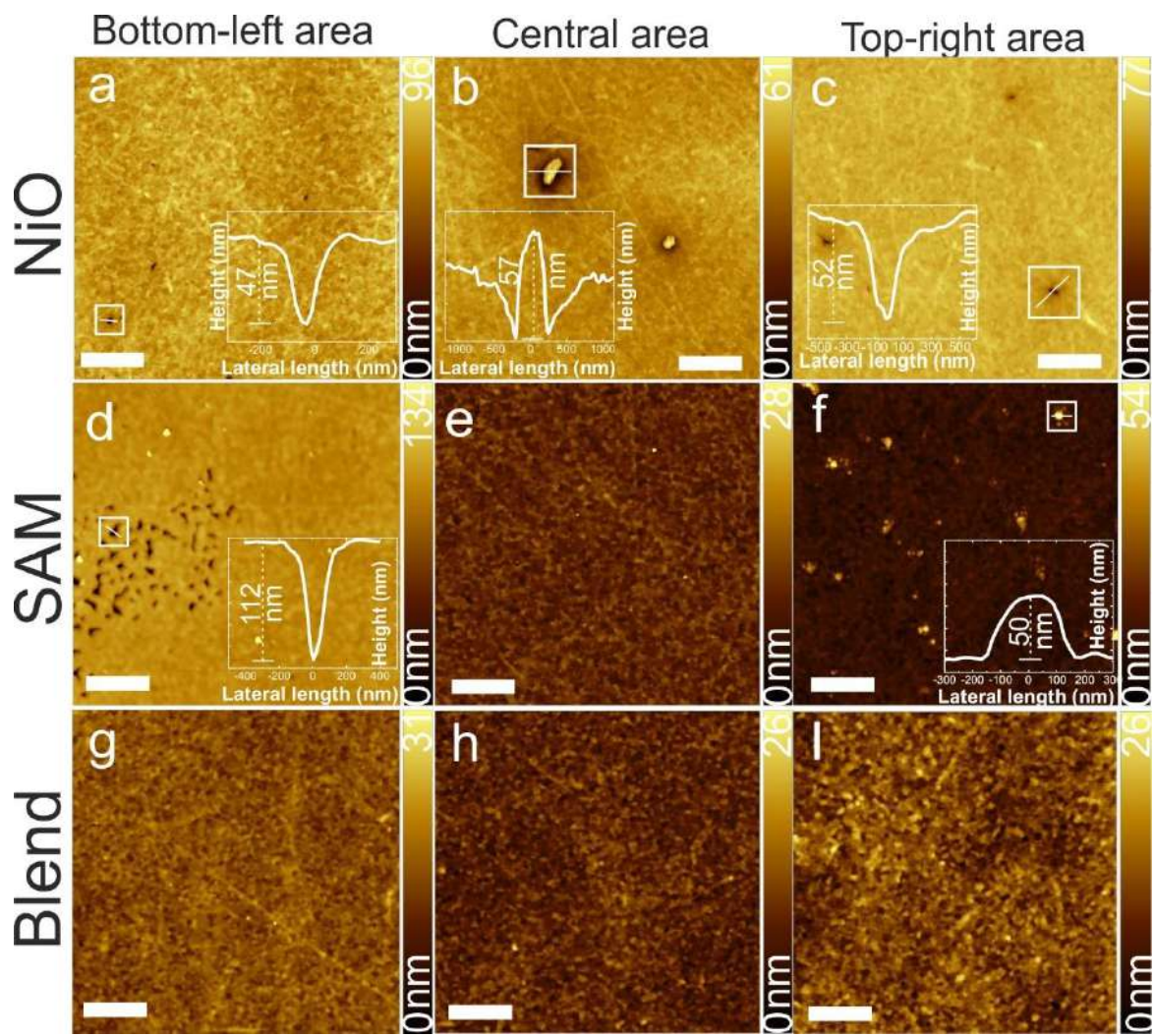

Figure 4 – The morphology of perovskite buried interfaces, crystallized on substrate location with different CWA. SAM BL (a), M(b), TR(c) and blend BL(d), M (e), TR(f), scale bar 2 μm

Changes in the morphology of buried interfaces for the perovskite absorber with a SAM interlayer resulted in significant fluctuations in the film thickness on the substrate (see box-charts in **Fig.S12** and **tab.S1** in **ESI**). The perovskite layers on the NiO and Blend surfaces had average thicknesses of approximately 540 nm with a standard deviation of 25 nm. The sample of SAM configuration exhibited variations in layer thickness, ranging from 300 to 550 nm, with an average standard deviation of 85 nm. This resulted in inhomogeneous absorption in different regions of the substrate with slot-die-coated HTLs and absorbers (corresponding spectra presented in the **fig.S13** in **ESI**). X-ray phase analysis of the three regions on SAM and Blend substrates is shown in **Figure S14** in the **ESI**. In all samples, the P4/mbm tetragonal phase of β-CsFAPbI$_3$ predominates, with peaks at 14.08º, 19.96º, 22.33º, 24.50º, 26.51º, 28.35º, and 31.80º. In the top-right location on the Blend and SAM substrates, low-intensity peaks corresponding to planes 001 and 101 of the hexagonal phase P-3m1 (164) of lead iodide were detected. **Table S2 (ESI)** presents the lattice parameters for all sample types, calculated using the Rietveld method in the Fullprof [35]. The lattice parameter $c$ decreases for all films relative to the reference, potentially indicating the presence of compressive stresses in the perovskite film formed during crystallization on organic interlayers. A decrease in the parameter $a$ a is characteristic of samples on substrates with a Blend interlayer. For both substrates, there is a common pattern: the maximum compressive stress appears in the center of the substrate. After a comprehensive analysis of the optoelectronic, structural, and surface properties of thin-film stacks with organic interlayers, we proceeded to characterize the devices.

We investigated p-i-n oriented PSMs with in-series connected 12 sub-cells and a total active area of 14.9 cm$^2$. Champion output performance of the PSMs measured under standard conditions presented in the **fig. 5(a)** and **tab.1**. The average open-circuit voltage ($V_{oc}$) values increased from 11.72 V for NiO modules to 11.88 V for SAM and 12.08 V for the Blend configurations, respectively (box-charts with statistical distribution of the data presented in **Fig.S15(ESI)**). Notably, the champion $V_{oc}$ values of SAM modules surpassed that of the Blend and didn't fully correlate with the average data. This underscores the inconsistent reproducibility of performance for SAM modules in batch fabrication experiments using slot-die coating technique. Modifying NiO PSMs with organic interlayers increased the short-circuit current ($I_{sc}$) values from 23.63 to 24.78 mA for blend modules (+4.8% enhancement). The SAM devices exhibited a negligible increase in the average photocurrent to 23.97 mA. The NiO PSM modules had an average fill factor (FF) of 66%, while the Blend configuration modules had a slightly improved value of 68%. The best modules from both configurations displayed a relevant FF of 74-75%.

In contrast, the SAM modules showed the lowest average fill factor of 62%, with the champion value of 66%. The IV curve analysis showed that SAM devices exhibited increased ohmic losses. NiO and Blend modules exhibited nearly equal series resistance ($R_s$), ranging from 1.98 to 2.05 Ohm. The SAM PSM displayed a significantly higher resistance of approximately 3.4 Ohm (> 34% larger). As result, the power conversion efficiency of the best (average) PSMs in Blend configuration reached 15.83% (13.74%), for NiO configuration - 14.63% (12.30%), and for SAM – 13.98% (11.79%). Despite the increase in $V_{oc}$ and $J_{sc}$, the performance of SAM PSMs was affected by a reduction in FF. Ohmic losses can originate from a localized inhomogeneity of TPATC interlayer and the heterogeneous properties of the perovskite absorber crystallized on its surface. TPATC clusters may negatively affect conductivity and introduce additional resistance at the hole transport interface. The Blend configuration effectively addresses the issue of inhomogeneities in the slot-die coated interlayer and absorber over a large area. This results in the desired key-properties of SAM for surface passivation and energy level alignment at hole collection junction in up-scaled PSMs. Stability analysis of the PSMs at the maximum power point ($P_{max}$, **fig.5(b)**) demonstrated a stable output with fluctuations ranging from 0.5% to 1.5% for all fabricated configurations.

The external quantum efficiency spectra for small-area single cells (see experimental for the details) presented in the **fig.5(c)**. For all devices with organic interlayers, we observed a decrease in the EQE in the short-wavelength region (320-430 nm), which is related to their parasitic absorption. For the SAM and Blend configurations, the EQE reduction was moderate, ranging from 5% to 8% compared to the control NiO. In the wavelength range of 420-620 nm, the highest EQE value was obtained for the Blend sample, reaching approximately 90% at 580 nm. The SAM sample achieved around 88%, while both the NiO and polymer samples showed EQE values of about 85%. Notably, the major difference in the EQE spectrum type between devices with organic interlayers and the control NiO was observed in the near-infrared region (620-780 nm). The NiO device exhibited a decrease in EQE down to 70%, whereas the configurations with organic interlayers maintained the long-wavelength shoulder EQE at a relatively high level of over 80%. For AM 1.5 G illumination conditions, the calculated $J_{sc}$ values, derived from integrating the EQE spectra, were: $J_{sc\ NiO}$ = 20.35 mA/cm$^2$; $J_{sc\ SAM}$ = 21.44 mA/cm$^2$ and $J_{sc\ Blend}$ = 22.10 mA/cm$^2$. These results are consistent with the IV measurements. In p-i-n architectures of PSCs, low-energy photon absorption and electron collection take place on the back side of the perovskite absorber layer. The observed reduction in EQE for NiO devices in the near-infrared region suggests reduced efficiency in electron transport within this solar cell configuration. Despite the optical losses in the 320-430 nm region, the use of organic interlayers improved photocarrier collection efficiency. Specifically, the increased EQE in the long-wavelength region can be attributed to the reduced impact of the electron traps in devices with organic interlayers. This may result from the improved quality of microcrystalline perovskites formed on the modified surface compared to NiO. The transport properties of the devices were investigated using dark IV curves (**Fig. 5(d)**). All configurations exhibited similar rectifying properties and no hysteresis. Notably, the use of organic interlayers of all types reduced the dark leakage current by almost an order of magnitude, from approximately $10^{-7}$ A/cm$^2$ (NiO) to $10^{-8}$ A/cm$^2$ (SAM and Blend). This indicates a reduced impact of micro-pinholes and shunting.

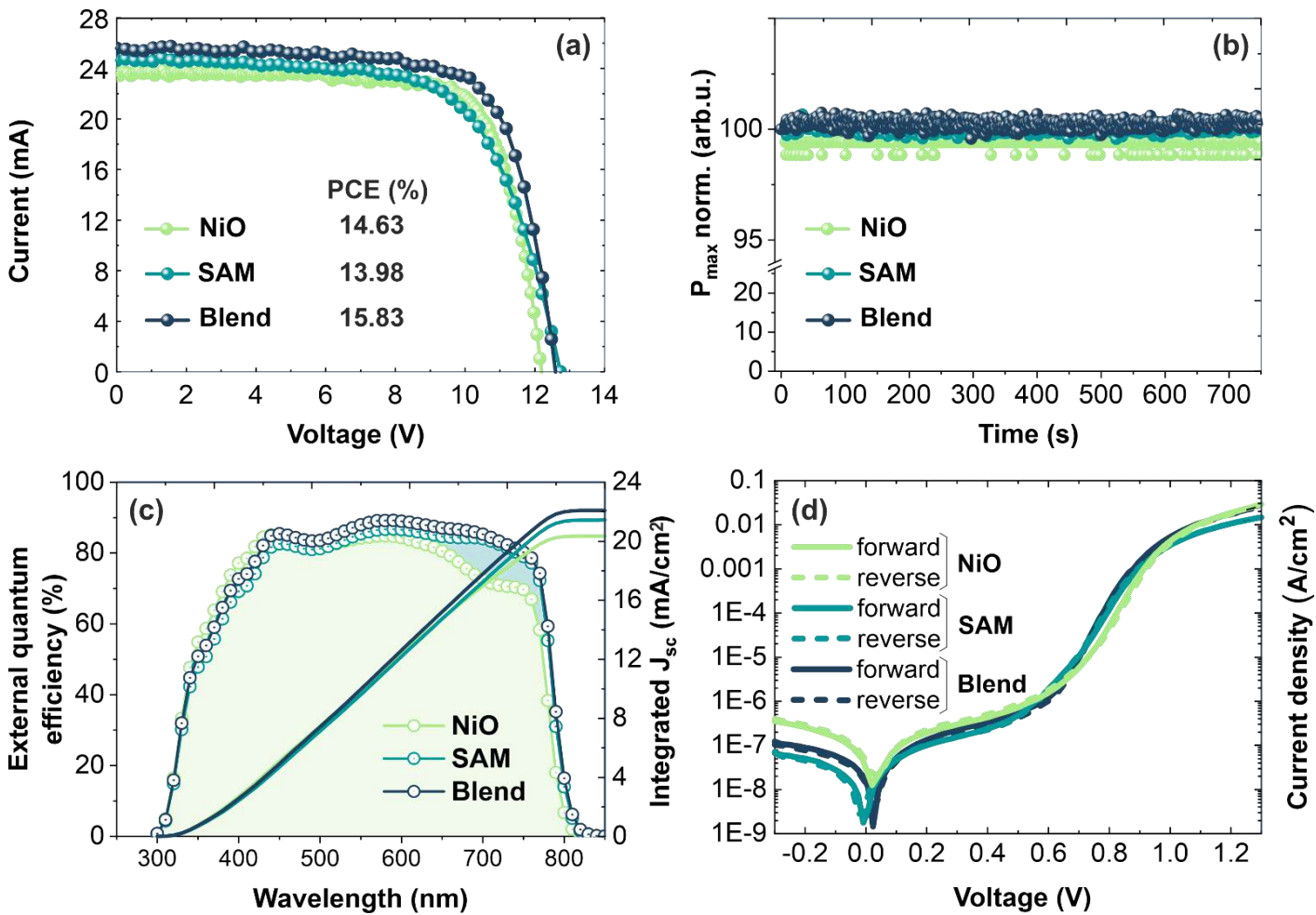

**Figure 5** – The champion IV curves for slot-die coated PSMs(a); stabilization of $P_{max}$ for PSMs (b); EQE spectra for the small area devices of different configurations (c); dark JV curves for small-area device fabricated with different organic interlayers

**Table 1.** The parameters of the output IV performance for slot-die coated PSMs

| PSM configuration | $V_{oc}$ (V) (aver. value) | $I_{sc}$ (mA) (aver. value) | FF (aver. value) | PCE (%) (aver. value) | $P_{max}$ (mW) (aver. value) |
|---|---|---|---|---|---|
| NiO | 12.23 (11.72) | 23.63 (23.68) | 0.75 (0.66) | 14.63 (12.30) | 217.69 (183.03) |
| SAM | 12.73 (11.88) | 24.73 (23.97) | 0.66 (0.62) | 13.98 (11.79) | 208.02 (175.79) |
| Blend | 12.48 (12.08) | 25.58 (24.78) | 0.74 (0.68) | 15.83 (13.74) | 235.55 (204.50) |

The charge-carrier extraction properties were investigated through transient photocurrent (**TPC**) measurements (**Fig. S16** in **ESI**) for small-area devices. When a solar cell is illuminated with a short pulse of light, the photo-generated charges are extracted at the electrodes, resulting in a measured photocurrent. This process provides valuable insights into the dynamics of charge extraction and recombination processes within the device. The obtained TPC data represents the rising and falling profiles of the photocurrent under short-circuit conditions. Rise time ($t_r$) is defined as the period required for the photocurrent to increase from 10% to 90% of its saturated value during the light-on process. Similarly, the fall time ($t_f$) is determined is calculated in the opposite direction. The rise time for NiO devices was 2.89 μs, while the SAM and Blend configurations exhibited more rapid dynamics, with $t_r$ values of 2.32 μs and 2.23 μs, respectively (**fig.S16(a) in ESI**). For measurements in the FALL mode (**Fig.S16(b) in ESI**), PSCs with organic interlayers exhibited a 20-25% faster response time compared to the unmodified NiO sample. Decrease in rise and fall times for PSCs with SAM and blended organic interlayers indicated more efficient charge extraction and suppressed impact of the non-radiative recombination properties. The rapid filling of trap states during photo-injection suggests a reduction in defect concentration at the hole-transport interface. The obtained data is consistent with morphological analyses of the buried hole transport junction and photo-carriers' lifetime measurements extracted from time-resolved PL measurements (TRPL). The decay of both components ($\tau_1$ and $\tau_2$) in the photoluminescence of perovskite films increased from NiO to SAM and Blend (Table 2). This can be explained by a more efficient hole-selective contact and altered the interfacial charge transfer dynamics, resulting in faster charge extraction for a SAM and Blend samples[36].

**Table 2.** TRPL normalized data for perovskite films fabricated on NiO; NiO/TPATC; NiO/TPATC/polymer; NiO/polymer

| Sample configuration | | $A_1$, ÷10³ | $\tau_1$ (ns) | $A_2$, ÷10³ | $\tau_2$ (ns) | $\tau_{average}$ (ns) |
|---|---|---|---|---|---|---|
| **NiO** | median | 16.6 | 8.1 | 9.6 | 41.4 | 31.1 |
| | average | 16.5 | 8.2 | 9.3 | 41.6 | 32.9 |
| | ± std | ± 3.5 | ± 0.4 | ± 2.1 | ± 4.8 | ± 4.8 |
| **SAM** | median | 14.4 | 8.6 | 12.3 | 43.4 | 36.0 |
| | average | 14.3 | 8.8 | 12.2 | 42.8 | 36.2 |
| | ± std | ± 2.2 | ± 0.5 | ± 1.6 | ± 3.4 | ± 3.2 |
| **Blend** | median | 15.1 | 9.5 | 15.1 | 57.5 | 50.7 |
| | average | 15.2 | 9.8 | 15.4 | 56.9 | 50.0 |
| | ± std | ± 0.4 | ± 1.1 | ± 2.0 | ± 4.9 | ± 5.5 |

*Notes: A1 and A2 is intensity amplitudes; $\tau_1$ and $\tau_2$ is luminescence lifetimes; $\tau_{average}$ is a weighted average luminescence lifetime; std is standard deviation.*

Following the analysis of the performance of PSMs under standard conditions, we measured the temperature coefficients for output performance and IV measurements under ambient conditions with low-light illumination. Variations in temperature and non-standard lighting are expected factors for real exploitation of solar panel. Temperature coefficients for IV parameters are typically documented in solar panel data sheets. A key indicator of changes in the energy yield of a solar cell, when the temperature deviates from the standard value, is calculated according to **eq. 1**[37].

$$T_{Pmax} = \frac{1 - norm.Pmax_{AT}}{T_{AT} - T_{RT}} \quad (1)$$

Where 1- normalized $P_{max}$ at standard conditions;

AT- actual temperature;

RT – room temperature (25°C);

To assess the performance of PSMs at various temperatures, we utilized a thermostatic bench equipped with Peltier elements. During the study, we analyzed the variations in IV parameters across a temperature range from -5°C to 35°C under illumination from AAA solar simulator with an AM 1.5 G light spectra. We performed two types of measurements, which included gradual cooling and heating. For the first case (cooling), the PSM was measured at room temperature, then heated to 35°C, after which we performed cooling measurements in 5°C increments. The second type of measurements (heating) proceeded in the reverse order. After measurements at RT, we cooled the device to −5°C and performed gradual heating. The heating/cooling rate was of the order of 1°C per minute, and we kept the temperature constant for at least 1 minute before measuring the IV curves. The analysis of the calculated energy yield for PSMs revealed significant differences between configurations with various organic interlayers (**fig.6** for cooling mode and **fig.S17** in **ESI**). We note that the measurements in the heating and cooling regimes correlated strongly near 25°C, exhibiting similar trends but lacking full repeatability at the temperature extremes of -5°C and +35°C. Generally, we observed an increase in $V_{oc}$, $J_{sc}$, and $P_{max}$ when the samples were cooled to -5°C, while at higher temperatures (up to 35°C), these characteristics slightly decreased. $T_{Voc}$ for NiO PSMs reached rel. +0.07%/°C at maximum cooling. The polymer and blend configurations showed the same value of $V_{oc}$ change dynamics +0.14%/°C(**fig.6(a)**). For the SAM module, the $T_{Voc}$ at reduced temperature increased more than 3 times to +0.26%/°C relative to NiO devices. For the elevated temperature, control PSMs without organic interlayer showed the value of −0.22%/°C. In contrast, at °C SAM, and blend devices didn't exhibit any meaningful changes of $V_{oc}$ with respect to values at 25°C. The trend of $J_{sc}$ versus temperature(**fig.6(b)**) exhibited a close correlation for PSMs during cooling, with values ranging from +0.16 to +0.18%/°C for NiO, and blend devices, and an increase up to +0.21%/°C for SAM. The behavior of the photocurrent at higher temperatures showed the largest decrease for NiO at −0.18%/°C, and an improvement of approximately to −0.12%/°C for SAM and Blend. The dependence of FF on device temperature (**fig.6(c)**) showed overlapping data and was difficult to characterize, although a negative trend was identifiable for all configurations with organic interlayers. $T_{Pmax}$ analysis(**fig.6(d)**) shows the general dynamics of changes in device performance, which was characterized by a significant increase in performance for modules at low temperatures and a decrease in $P_{max}$ at elevated temperatures. $T_{Pmax}$ values for PSMs presented in the **tab.3**.

Table 3. Calculated $T_{Pmax}$ values for various configurations of slot-die coated PSMs

| $T_{Pmax}$ (%/°C) | NiO | SAM | Blend |
|---|---|---|---|
| Low temperature (-5°C) | +0.19 | +0.37 | +0.40 |
| Elevated temperature (+35°C) | -0.34 | -0.53 | -0.39 |

For the case of analyzing $T_{Pmax}$ changes in a linear approximation, we calculated the slope coefficients (**K**) of the corresponding fitting lines. Modules with an organic interlayer containing SAM exhibited the steepest slope for the $T_{Pmax}$ dependence, with values of $K_{SAM}$ =0.0041 and $K_{Blend}$=0.0042. For the PSM with NiO configuration, the slope was significantly smaller, $K_{NiO}$ =0.00267. The temperature-dependent behavior of $P_{max}$ for slot-die coated PSMs revealed high sensitivity to the surface type at the HTL/perovskite interface.

So, each modification of the HTL/perovskite has a unique impact on the device operation at different temperatures. Current literature highlights[38–40] that temperature effects in perovskite solar cells exhibit a complex dependence on changes in the absorber bandgap, surface recombination dynamics, and variations in energy level alignment. Spectral changes in the perovskite layer properties typically affect $J_{sc}$ predominantly. In our study, we observed no significant changes in the band-gap of the perovskite absorber across different surface types, and the $T_{Jsc}$ values for PSMs with various configurations were closely aligned. Our data suggest that variations in $V_{oc}$ are primarily responsible for changes in $T_{Pmax}$. According to the diode equation analysis, $V_{oc}$ depends on parameters such as charge carrier lifetimes, intrinsic carrier concentration, non-ideality factor, and others. Temperature increases result in higher intrinsic carrier concentrations, leading to increased dark saturation current and reduced $V_{oc}$. Charged defects in the microcrystalline perovskite layer can significantly impact carrier transport and concentration, including inducing accumulation processes. The HTL interface state can strongly affect defect types, their thermal activation thresholds, and potential ion-electron interactions. To investigate the deep insights into low-temperature PSCs' operation with respect to $V_{OC}$ temperature dependence, we employed Photo-Induced Voltage Transient Spectroscopy (PIVTS)[41] and Admittance Spectroscopy (AS) for small-area devices.

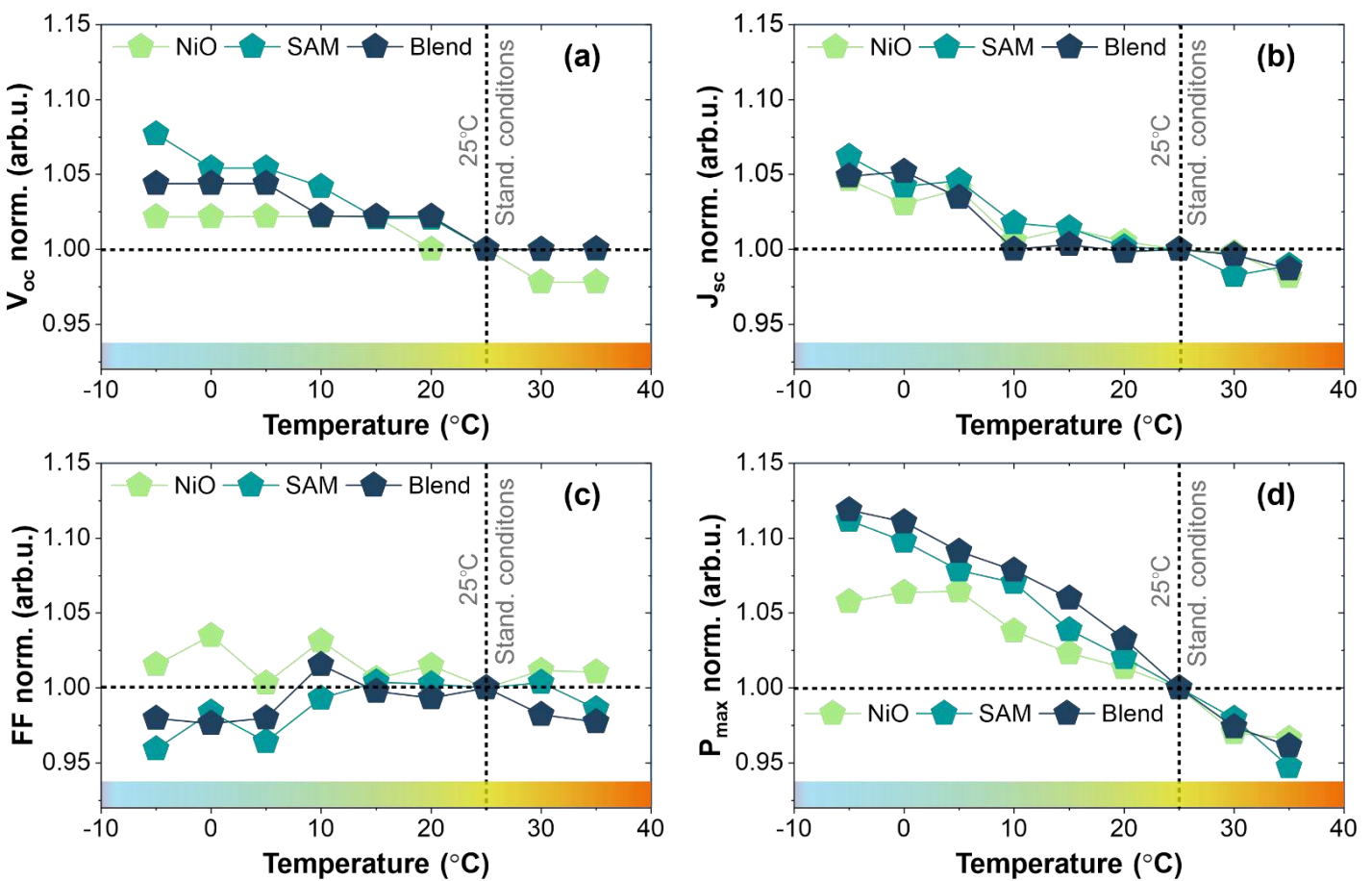

**Figure 5** – The temperature dependence of the IV parameters for slot-die coated PSMs under AM 1.5 G light (100 mW/cm$^2$). $V_{oc}$ vs temperature (a), $J_{sc}$ vs temperature (b), FF vs temperature (c), $P_{max}$ vs temperature (d)

The concept behind PIVTS is to track the open-circuit voltage ($V_{OC}$) of the device after a light pulse at different temperatures to estimate the $V_{OC}$ relaxation kinetics. When illuminated (in this study, we used a 470 nm LED), the solar cell reaches its steady-state $V_{OC}$ value. After turning off the light, all non-equilibrium carriers recombine within their lifetime (typically in the nanosecond range), leading to rapid $V_{OC}$ relaxation. However, in perovskite solar cells, this process is slowed down by mobile ions, which participate in ionic currents[42,43] and screen electric fields[44]. Therefore, capturing the time dependence of $V_{OC}$ after a light pulse can provide insights into mobile ion kinetics and allow us to estimate $V_{OC}$ within a specific temperature interval. Therefore, capturing the time dependence of $V_{OC}$ after a light pulse can provide insights into mobile ion kinetics and allow us to estimate $V_{OC}$ within a specific temperature interval. PIVTS measurements plots presented in the **fig.6 (a)-(c)**, the AS data is shown in the **fig.6(d)**. The temperature dependence for $V_{oc}$ of PSCs illuminated with LED presented in the **fig.6(e)**.

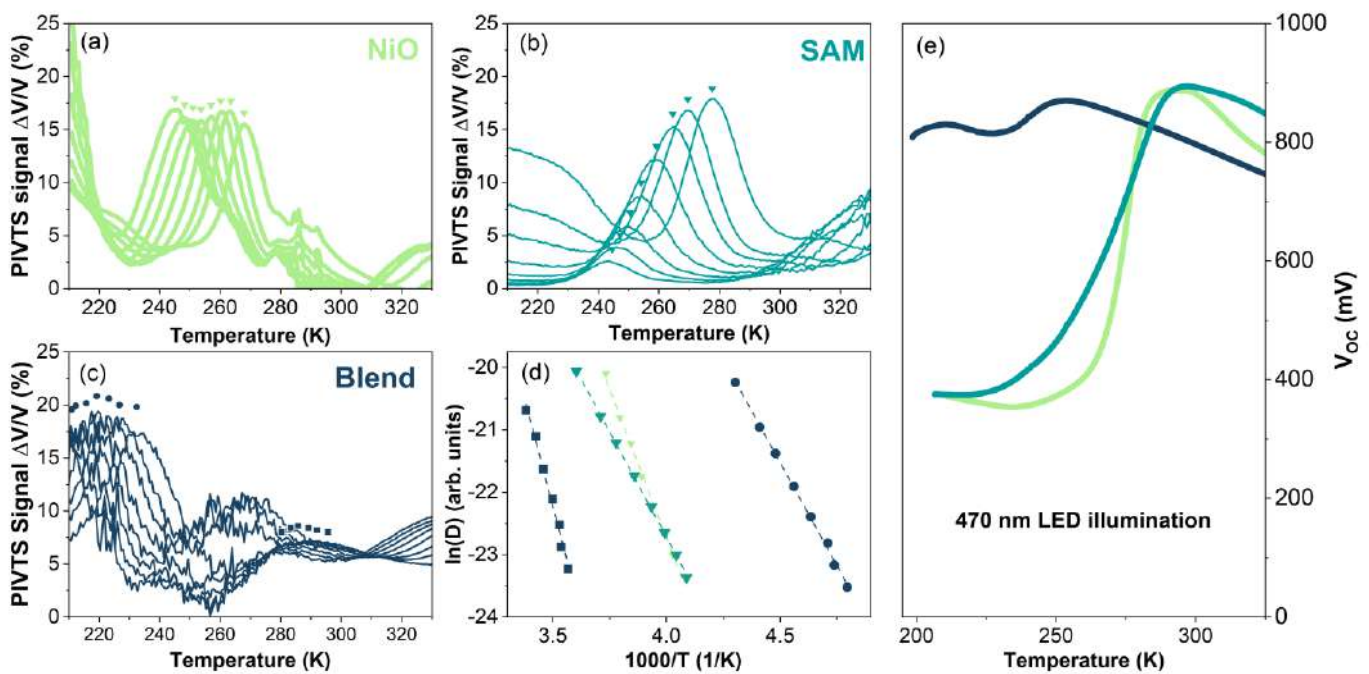

**Figure 6** - PIVTS measurements for NiO(a), SAM(b) and Blend(c) PSCs; admittance spectra (d) and $V_{oc}$ vs. temperature dependence for small-are devices

The results obtained for NiO and Self-Assembled Monolayer (SAM) samples exhibited similar trends in many aspects. Both samples demonstrated a temperature dependence of the steady-state open-circuit voltage ($V_{OC}$) that doubled upon heating above 250 K, reaching a maximum at 288-295 K followed by a linear decrease. The Blend sample exhibited much more stable $V_{OC}$ temperature dependence attributed to its high-quality electrical contact. The $V_{OC}$ remained around 800 mV across a wide temperature range with small fluctuations potentially caused by mobile ion contributions to the PIVTS signal. We observed mobile ion contributions to the PIVTS signal in both NiO and SAM samples within a similar temperature range around 260 K. In contrast, the blended sample demonstrated the presence of two mobile ion species at 220 K and 290 K. These mobile ion features could be associated with energy activation data from the actual literature[45–50]. It is important to note that the similar activation energy values and numerous technological factors make it challenging to precisely identify the defect types. In this study, we differentiated defect centers by their characteristic temperature ranges of activity. For NiO and SAM device configurations within the 240-280 K temperature range (PIVTS), we identified defect centers **α**, with activation energies of 0.79 and 0.57 eV, respectively. Potentially, the states of these centers may correspond to defects related to the iodine anion, such as iodine vacancies and iodine interstitials. The Blend sample showed a similar activation energy value but at a higher temperature, indicating a different nature for this defect (probably a formamidinium vacancy, assigned to state **β**). A characteristic feature of Blend devices was the presence of a deep center **δ** with an activation energy of 1.2 eV, which appeared at low temperatures around 220 K. Identifying these states is quite complex. However, a general analysis of the PIVTS spectra indicates that the Blend configuration exhibits a reduced signal amplitude at lower temperatures of 240-290 K and a reconfiguration of states compared to NiO and SAM. AS measurements revealed close activation energy values ranging from 0.56 to 0.61 eV for all device types (**Fig. S18** in **ESI**). Combining AS and PIVTS data across the entire temperature range showed a strong correlation (linear trend) for α and β group defects (**Fig. S19** in **ESI**). The summarized data from AS and PIVTS measurements presented in the **tab.4.**

Table 4. Summarized data for extracted energy activation values from PIVTS and AS spectra

| Measurement technique | NiO (defect type) | SAM (defect type) | Blend (defect type) |
|---|---|---|---|
| PIVTS | 0.79 (α) | 0.57 (α) | 0.56(β) / 1.20 (δ) |
| AS | 0.57 (α) | 0.56 (α) | 0.61 (β) |

After a comprehensive analysis of the photovoltaic parameters for slot-die coated PSMs, we analyzed the stability of their performance under continuous photo-stress (**fig.6**). We evaluated the relative change in $P_{max}$ under $V_{oc}$ conditions at a temperature of 63.5±1.5°C, in accordance with the ISOS-L-2 protocol[51]. For all types of modules, we observed a burn-in period during the first 100 hours, after which the maximum power showed a negative trend. Stability testing of PSMs was performed during $T_{80}$ period—time when the device loses 20% of its initial $P_{max}$ value. The control NiO PSM demonstrated a relevant $T_{80}$ of approximately 1000 hours, while the SAM configuration didn't reach the 500-hour period. In contrast, the Blend PSM exhibited exceptional stability, with a $T_{80}$ of 1630 hours. We assume that the decreased stability of SAM configuration samples is mainly due to the presence of inhomogeneous clusters. The formation of macro-defects at the hole-transport interface and the reconfiguration of ionic defects states can trigger the decomposition of perovskite, leading to the formation of non-radiative centers.

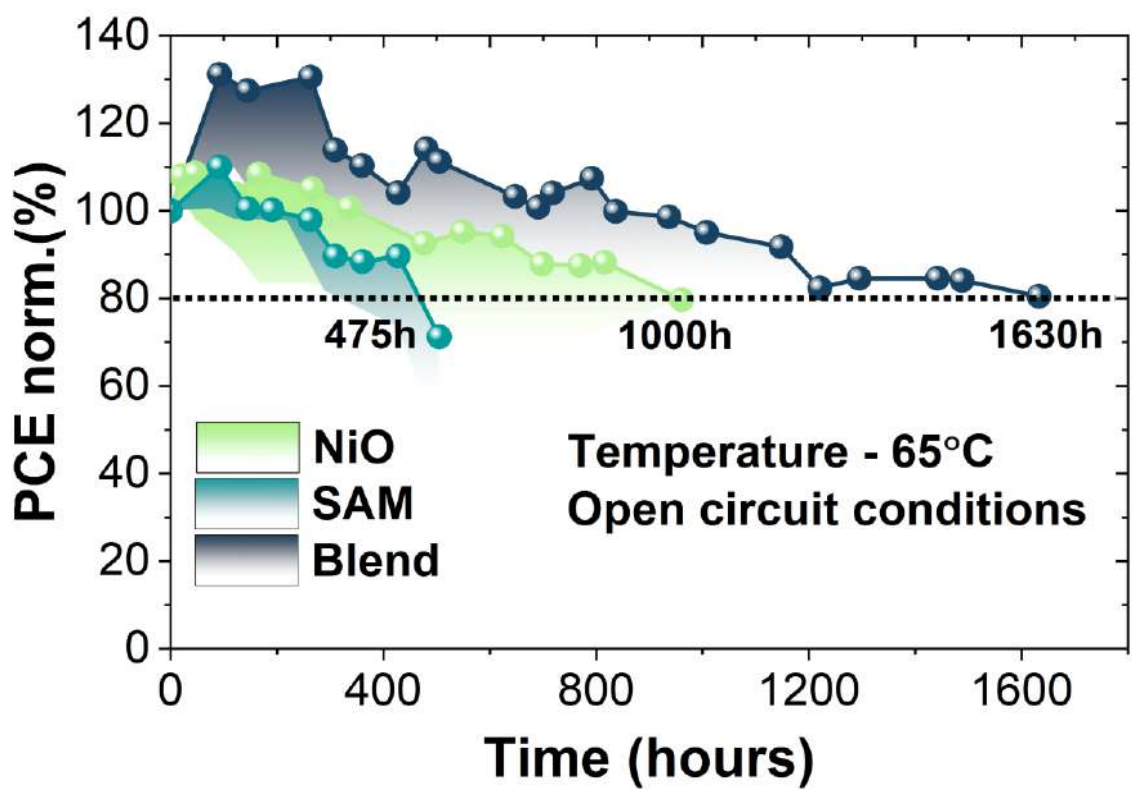

**Figure 7**– The stability performance of slot-die coated PSMs under continuous light soling and temperature of 65°C. The solid line with points represents the best-performing modules, while semitransparent areas demonstrate the data distribution for the samples from the batch

The influence of traps on solar cell performance is more pronounced under low-light conditions, when the concentration of photo-injected carriers significantly decreases compared to AM 1.5 G illumination. Non-radiative losses, shunting effects, and high-contact resistance can severely impact the performance of PSMs in cloudy weather, shading, and other external conditions. To assess the impact of the developed organic interlayers in slot-die coated PSMs, we measured IV performance under ambient conditions in the Moscow region during cloudy weather (May 2024, **fig. S20(a,b)** in **ESI**). The light intensity range (wavelengths 500-700 nm) varied from approximately 100 to 15,000 lux at reduced air temperatures (11.8 to 13.6°C). Under various ambient conditions, the performance of PSMs was primarily characterized by the $V_{oc}$ dependence on light intensity (**Fig. S20(c)** in **ESI**), leading to changes in Pmax (**Fig. S20(d)** in **ESI**). At higher light intensity >10, 000 Lx the $V_{oc}$ values followed the trend of the data for standard conditions. Blend and SAM PSMs reached the $V_{oc}$ close to 12 V, while NiO device exhibited the reduced value of ~11.2 V. At light intensities ($I_L$) below 8,000 lux, the dynamics of $V_{oc}$ decrease relative to $I_L$ became significantly steeper for the SAM module. Thus, at a minimum illumination of 70-90 lux, $V_{oc}$ for Blend was 7.2 V, $V_{oc}$ for NiO was 5.9 V, and for SAM was 3.3 V. This resulted in a significant difference in the generated maximum power for the modules under low-light condition, where Blend configuration demonstrated superior performance compared to analogues.

The obtained stability and performance data for perovskite solar modules align with the current state-of-the-art levels and are consistent with existing literature[52–55]. Scaling up solution-processing methods involves several factors, including the use of surfactants[56,57], solvent mixture engineering[58], and other techniques. It is widely recognized that the surface condition has a direct effect on the crystallization processes of perovskite absorbers. Many studies[59–61] report that an increase in the contact wetting angle leads to a reduction in the number of nuclei during liquid-phase crystallization, thereby increasing the average grain size and reducing intergranular boundaries. The practical application of slot-die layer-by-layer deposition demonstrates the need for precise adjustment of surface hydrophobicity. Thus, our two-component interlayer approach provides advantages that can be widely applied to various technological processes. Selecting the appropriate types and ratios of polymers and SAMs) in the interlayer can balance surface wetting and energy level positions for a given perovskite composition. This approach can be applied not only to scaling up solution processing but also to vacuum deposition of perovskite films (thermal evaporation), which is extremely sensitive to the crystallization surface state. The incorporation of polymer in the blend interlayer not only enhances homogeneity but also reduces the strain effects at the perovskite/HTL contact, serving as a relaxing buffer. This development enables a more efficient realization of SAM's potential for interface modification and addresses associated technological complexities.

**Conclusions:**

In this study, we conducted a comprehensive investigation of the technological optimization of SAM slot-die coating for perovskite modules with NiO HTL. Traditional SAM solutions typically result in inhomogeneous coatings, which impair their functionality in slot-die applications and subsequent layer-on-layer perovskite deposition. Our findings indicate that standard wet film formation followed by vacuum treatment results in SAM-clusters with very high CWA up to 90°. This leads to unfavorable crystallization conditions for the perovskite absorber, causing macro-defects (~100 nm voids, ~1 μm disoriented crystallites) on the buried interface, significantly affecting the coating's thickness and uniformity. The presence of inhomogeneous regions substrate affected contact properties of the PSMs with SAM. The PCE of SAM modules was affected by with best(average) values of 13.98 (11.79) %, which was smaller compared to the NiO 14.63 (12.30) %. To address the wettability issue, a blended solution comprising a pTPA-TDP polymer was developed, with the objective of optimizing viscosity and decreasing the CWA (46°) of the surface. This resulted in enhanced buried interface quality and interlayer morphology, which in turn led to a gain of PCE of up to 15.83%, primarily due to the improved $V_{oc}$ and $J_{sc}$, observed in the SAM-based device, with the additional benefit of enhanced contact properties. Also, by applying the blend interlayer, the parameters of defect centers were reconfigured. Defect center spectroscopy revealed that the 0.5-0.8 eV deep states observed in all configurations exhibit different temperature ranges, which strongly affect the $V_{oc}$ when the devices are cooled.

To the best of our knowledge, this work is the first to report a successful adaptation of SAM interlayers for slot-die coating cycle of perovskite modules. Our novel approach, which allowed us to tailor SAM properties for slot-die coating without altering its synthesis. Notably, we verified our method on modules with slot-die coated HTLs and absorbers, whose fabrication is most critical for layer-on-layer deposition. The results demonstrate the importance of not only tuning the semiconductor properties but also the technical implementation of the approach for technology scaling. Changing the properties of passivation layers has a complex effect on device performance and the nature of defects. Improving the perfection of thin-film structures with the blend interlayer significantly increased the stability of modules up to 1630 hours. We believe that further progress in developing the slot-die coating cycle for perovskite photovoltaics will focus on adapting double-sided passivation techniques using SAM for the electron collection interface. Unifying process steps for passivation interlayers will bring the performance of PSMs closer to industry standards, similar to hetero-structured Si cells with amorphous layers.

**Acknowledgments**

The work was supported by the Russian Science Foundation (project № 22-19-00812) - https://rscf.ru/project/22-19-00812/. NMR spectra were recorded using the equipment of the Collaborative Access Center 'Center for Polymer Research' of ISPM RAS with support from the Ministry of Science and Higher Education of the Russian Federation (topic FFSM-2024-0003).

"Tailoring wetting properties of hole-transport interlayers for slot-die coated perovskite solar modules"

By

T.S. Le [a],[1] I.A. Chuiko[a,b],[1] L.O. Luchnikov[a], K.A. Ilicheva[a], P.O. Sukhorukova [a,b], D.O. Balakirev[b], N.S. Saratovsky [b], A.O. Alekseev[a], S.S. Kozlov[c], D.S. Muratov[d], V.V. Voronov[a], P.A. Gostishchev[a], D. A. Kiselev[e], T.S. Ilina[e], A.A. Vasilev[f], A. Y. Polyakov[f], E.A. Svidchenko,[g] O.A. Maloshitskaya[g], Yu. N. Luponosov [b]* and D.S. Saranin [a]*

[a]LASE – Laboratory of Advanced Solar Energy, National University of Science and Technology "MISiS", Leninsky Prospect 4, 119049, Moscow, Russia

[b]Enikolopov Institute of Synthetic Polymeric Materials of the Russian Academy of Sciences (ISPM RAS), Profsoyuznaya St. 70, Moscow, 117393, Russia

[c]Laboratory of Solar Photoconverters, Emanuel Institute of Biochemical Physics, Russian Academy of Sciences, 119334 Moscow, Russia

[d]Department of Chemistry, University of Turin, 10125, Turin, Italy

[e]Laboratory of Physics of Oxide Ferroelectrics, Department of Materials Science of Semiconductors and Dielectrics, National University of Science and Technology MISIS, Moscow 119049, Russia

[f]Department of Semiconductor Electronics and Semiconductor Physics, National University of Science & Technology MISIS, 4 Leninsky Ave., Moscow, 119049, Russia

[g]Moscow State University, Chemistry Department, 1/3 Leninskie Gory, Moscow, 119991, Russia

**Experimental details:**

## Materials for synthesis

2.5 M solution of *n*-butyllithium in hexane (*n*-BuLi), iron(III) chloride (Sigma-Aldrich Co), nitrobenzene (Pallav Chemicals) were used without further purification. Tetrahydrofuran (THF), toluene, ethanol and other solvents were purified and dried according to known methods. **TPA-TDP** was obtained as described in Ref.[1], 4,4,5,5-Tetramethyl-2-(2-thienyl)-1,3,2-dioxaborolane (**2**) was prepared as described elsewhere Ref.[2]. **TPA-T** was obtained as described in Ref. [3].

## Materials for perovskite absorber and charge-transporting layers

$NiCl_2·6H_2O$ (from ReaktivTorg 99þ% purity) used for HTM fabrication. Formamidinium iodide (FAI, >99.99%) was purchased from Greatcell Solar (Australia), Lead iodide ($PbI_2$, >99.9%) and cesium iodide (CsI, >99.99%) were purchased from Chemsynthesis (Russia) and LLC Lanhit (Russia), respectively. Fullerene-$C_{60}$ ($C_{60}$, >99.5%+) was purchased from MST (Russia). Bathocuproine (BCP, >99.8%) was purchased from Osilla Inc. (UK). Mxenes ($Ti_3C_2$) was purchased from Beijing Beike 2D materials (China). The organic solvents: 2-Methoxyethanol (2-ME), dimethylformamide (DMF), *N*-Methyl-2-pyrrolidone (NMP), chlorobenzene (CB), and isopropyl alcohol (IPA) were purchased in anhydrous from Sigma Aldrich and used as received without further purification.

Synthetic procedures

Synthesis of **pTPA-TDP** via oxidative polymerization has been described previously Ref.[4]. The oxidative coupling polymerization was carried out at room temperature in nitrobenzene solution using $FeCl_3$ as an oxidant. $FeCl_3$ (0.49 g, 3.01 mmol) and nitrobenzene (6 mL) were added to **TPA-TDP** (0.6 g, 1.21 mmol) in an inert atmosphere. The solution was stirred at room temperature for 240 h and poured into ethanol. The precipitate was collected and then the product was washed successively with 1M hydrochloric acid solution (3x3mL) and ammonia water (3x3mL). Purification was carried out by precipitation from a chloroform solution with ethanol and the Soxhlet extraction method in acetone, ethyl acetate, toluene, and ethanol. The resulting polymer was passed through a layer of silica gel (eluent - chloroform) to obtain a burgundy-colored solid (0.39 g, 65%). $^1$H NMR (250 MHz, $CDCl_3$): δ [ppm] 7.02-7.24 (overlapping peaks, 7H), 7.30 (s, 2H, $J$ = 7.33 Hz), 7.41-7.70 (overlapping peaks, 9H). $^{13}$C NMR (125 MHz, $CDCl_3$): δ [ppm] 75.47, 77.13, 114.18, 114.72, 115.92, 116.22, 122.15, 122.52, 123.73, 125.13, 125.30, 125.43, 125.58, 127.41, 127.67, 129.56, 131.66, 131.78, 131.88, 131.91, 135.85, 138.76, 145.62, 149.08, 156.01, 162.78, 166.10. Calcd (%) for $(C_{32}H_{18}F_1N_3S)_n$: C, 77.56; H, 3.66; S, 6.47; N, 8.48. Found (%) for $(C_{32}H_{18}F_1N_3S)_n$: C, 77.69; H, 3.86; S, 6.60; N, 8.37.

Molecular weight characteristics of the polymers were studied by gel permeation chromatography (GPC) using a polystyrene standard. Weight average molecular weight of the polymer was 11 000, number average molecular weight was 8 000.

Synthetic procedures of **TPATC** via Suzuki coupling reaction and carboxylation of the lithium derivative of product of previous reaction were described in our recent work[5]. The synthesis scheme is outlined in Fig. S1.

**5-[4-(Diphenylamino)phenyl]thiophene-2-carboxylic acid (TPATC).** *n*-BuLi (2.5 M solution in hexane, 2.4 mL, 6.1 mmol) was slowly added dropwise to a solution of **TPA-T** (compound **1**, 2.0 g, 6.1 mmol) in 40 mL of dry THF at –78°C. After complete addition, the reaction mass was stirred at –78°C for 1 hour. Thereafter, excess dried $CO_2$ gas was slowly bubbled through the stirring reaction mass for 2 hours at –70°C. After the end of the reaction, the cooling bath was removed, and the stirring continued for another 30 minutes with the temperature rising to RT. After completion of the reaction, the mixture was poured in 200 mL of diethyl ether, containing 15 mL of 1 M HCl solution. The organic phase was washed with water (250 ml), then the organic phase was separated, and the solvent was evaporated under vacuum. After purification by recrystallization from ethanol, **TPATC** (1.6 g, 70%) was obtained as a light green powder. M.p. = 238 °C. $^1$H NMR (250 MHz, $CDCl_3$, δ, ppm): 6.93 (d, 2H, $J$ = 8.80 Hz); 7.01–7.13 (overlapping peaks, 6H); 7.32 (t, 4H, $J$ = 8.26 Hz); 7.41 (d, 1H, $J$ = 3.85 Hz); 7.59 (d, 2H, $J$ = 8.80 Hz); 7.66 (d, 2H, $J$ = 3.85 Hz). $^{13}$C NMR (75 MHz, $CDCl_3$, δ, ppm): 122.32; 123.47; 123.88; 124.75; 126.32; 127.08; 129.79; 132.10; 134.54; 146.64; 147.96; 149.95; 162.94. IR ($cm^{-1}$): 1660 (C=O). Calcd. (%) for $C_{23}H_{17}NO_2S$: C, 74.37; H, 4.61; N, 3.77; O 8.61; S, 8.63. Found C, 74.53; H, 5.24; N, 3.61; S, 8.69. MALDI-TOF MS: found m/z 370.9; calcd. for [M]+ 371.1.

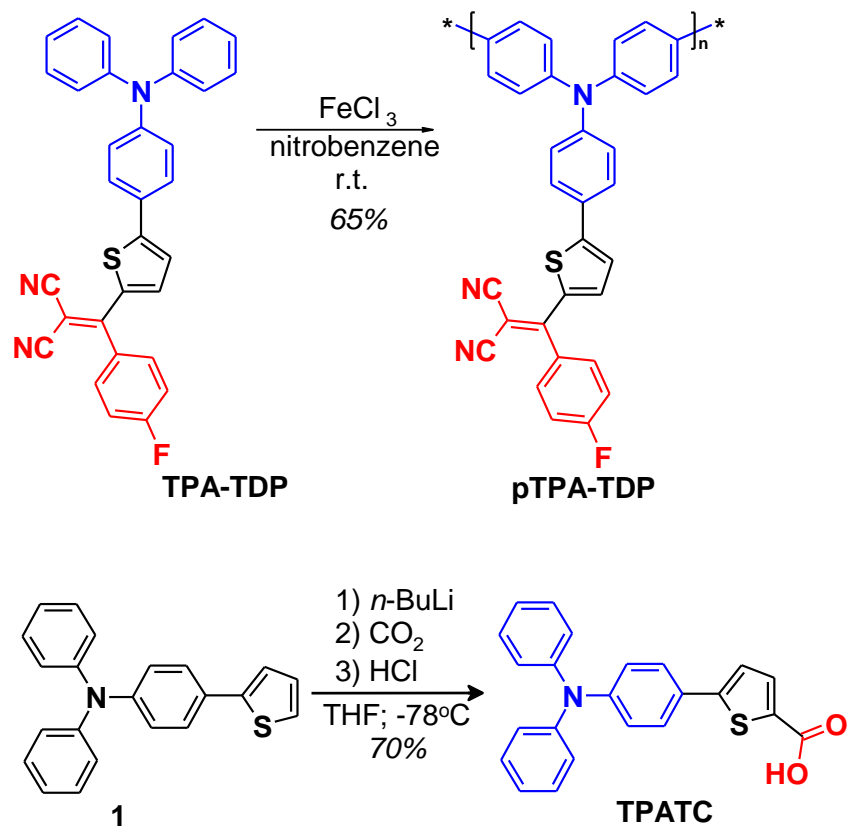

**Figure S1** – Synthesis of polymer **pTPA-TDP** and SAM **TPATC**

Substrates

Solar cells were fabricated on $In_2O_3$:$SnO_2$ (ITO) coated glass ($R_{sheet}$<7 Ohm/sq) from Zhuhai Kaivo company (China). Optical measurements were done with use of KU-1 Quartz substrates from LLC Alkor (Russia).

Preparation of the Precursors

Nickel oxide precursor ink was prepared by dissolving the $NiCl_2·6H_2O$ powder in 2-ME with a concentration of 50 mg/mL. Then the green solution was heated at 75 °C in 2 hours after adding nitric acid with a concentration of 25 μL/mL relative to 2-ME. The **TPATC** solution was prepared by dissolving 1 mg of **TPATC** in 2 mL of CB, and then the solution was heated at 50 °C in 1 hour before using. To prepare **TPATC:pTPA-TDP** solution we dissolving 1 mg of **TPATC** in 1 mL of CF and 1 mg of **pTPA-TDP** in 1 mL of CF in separate vials then heated at 50 °C in 1 hour. The two as-prepared solutions were mixed only before printing. For the perovskite precursor solution we mixed CsI, FACl, FAI and $PbI_2$ powders with stoichiometric molar ratio of 0.2:0.1:0.7:1 in DMF:NMP (vol. ratio 9:1) with a concentration of 0.6M, and then stirring at 60 °C for 30 min. Before printing the precursor, solution was filtered with 0.22 μm PVDF filters. For the hole-blocking layer, we used BCP-Mxenes solutions prepared by mixing 1 mg of BCP and 1 mg of Mxenes powders in 2 mL of IPA. The vial with solution was rinsed in ultrasonic bath for 15 min before stirring at 50 °C overnight to form the dispersion.

### The fabrication of 0.15 cm² active area slot-die coated PSCs

The planar inverted PSCs and PSMs have been fabricated with the following structures: glass/ITO/NiO$_x$/Cs$_{0.2}$FA$_{0.8}$PbI$_{2.7}$Cl$_{0.3}$/C60/BCP-Mxenes/Bi-Cu (NiO), glass/ITO/NiO$_x$/TPATC/Cs$_{0.2}$FA$_{0.8}$PbI$_{2.7}$Cl$_{0.3}$/C60/BCP-Mxenes/Bi-Cu (SAM) and glass/ITO/NiO$_x$/TPATC:pTPA-TDP/Cs$_{0.2}$FA$_{0.8}$PbI$_{2.7}$Cl$_{0.3}$/C60/BCP-Mxenes/Bi-Cu (Blend).

The ITO substrates were cut in a dimension of 25×25 cm² and patterned by UV-laser to isolate 5 mm wide semitransparent ITO electrodes as presented in [6]. Then the substrates were cleaned in an ultrasonic bath with detergent, de-ionized water, acetone and isopropyl alcohol (IPA) for 10 min each and UV/O3-treated for 30 minutes. The printing processes were performed on the slot-die coater (Ossila) with meniscus guide in air with RH of 20-40 % as described in the work[6]. The guide dimension (width×length) and the gap between substrate and meniscus guide were fixed at 25 mm × 500 μm and 150 μm respectively. NiO$_x$ precursor solution was printed on 80 ºC pre-heated substrate at 10 mm/s of coating velocity and 2 μL/s of solution feed rate. The as-printed substrates were placed on hot plate at 120 ºC and then annealed at 300 for 1 hour. The **TPATC** precursors were printed at printed at 25 mm/c of coating speed and 10 μL/s of syringe rate at room temperature. The as-printed substrates were immediately treated by vacuum assisted solution process (VASP) in 5 sec and annealed at 105 ºC in 5 min in air afterward. The perovskite precursor solution was coated at 28 mm/c of coating speed and 10 μL/s of syringe rate. Then the as-coated wet films were transferred into a vacuum chamber and kept in vacuum for 2 min. After that, the films were annealed in air at 105 ºC for 30 min. 30 nm C$_{60}$ based electron transporting layer was deposited with the thermal evaporation method at 10$^{-6}$ Torr vacuum level. The hole blocking layer BCP-Mxenes was spin-coated at 4000 RPMs and annealed at 50 °C (1 min). Finally, a 100 nm copper back electrode was deposited in a thermal evaporator through a shadow mask. The single cells were fabricated with 0.15 cm² active area measured equal to the intersection area of ITO and top Cu electrodes.

### 50×50 mm slot-die coated module fabrication

We fabricated PSMs with the standard in-series planar connection of 12 sub-cells using laser beam patterning, employing a pulsed nanosecond laser (355 nm, 5 W from LLC NordLase, Russia) according to the approach presented in [7]. P1 patterning was performed by UV-laser on 5×5 cm² ITO substrates to isolate semitransparent ITO electrodes. We used the same deposition procedures for functional layers of PSMs as described above with following corrected printing parameters to be compatible with larger substrate dimension: the guide dimension (width × length) was fixed at 50 mm × 500 μm; NiO$_x$ precursor solution was coated at 15 mm/s of coating velocity and 8 μL/s of solution feed rate; the TPATC precursors were printed at printed at 15 mm/c of coating speed and 10 μL/s of syringe rate and the coating speed and syringe rate for printing perovskite precursor were 15 mm/c and 12 μL/s respectively. P2 patterning was done by UV-laser before the deposition of 100 nm back electrodes (15 nm Bi and 85 nm Cu). The separation of metal electrodes (P3) was performed by UV-laser so that the dimension of single sub-cell is 4.00 × 0.31 cm² and the active area is 1.24 cm² respectively. That gives a module total active area of 14.88 cm². The

PSMs were encapsulated with UV-curable Epoxy (Osilla, UK) and cover glass to prevent interaction of materials with moisture and oxygen in ambient.

Characterization

**NMR spectra.** $^1$H NMR spectra were recorded using a "Bruker WP-250 SY" spectrometer, working at a frequency of 250 MHz and using CDCl$_3$ (7.25 ppm) or DMSO-d6 (2.50 ppm) signals as the internal standard. $^{13}$C NMR spectra were recorded using a "Bruker Avance II 300" spectrometer, working at a frequency of 75 MHz. In the case of $^1$H NMR spectroscopy, the compounds to be analyzed were taken in the form of 1% solutions in CDCl$_3$ or DMSO-d6. In the case of $^{13}$C NMR spectroscopy, the compounds to be analyzed were taken in the form of 5% solutions in CDCl$_3$ or DMSO-d6. The spectra were then processed on the computer using the "ACD Labs" software.

**Elemental analysis.** Elemental analysis of C, N and H elements was carried out using CHN automatic analyzer "CE 1106" (Italy). The settling titration using BaCl$_2$ was applied to analyze the S element.

**Mass-spectra.** Mass-spectra (MALDI-TOF) were registered on a "Autoflex II Bruker" (resolution FWHM 18000), equipped with a nitrogen laser (work wavelength 337 nm) and time-of-flight mass-detector working in the reflections mode. The accelerating voltage was 20 kV. Samples were applied to a polished stainless-steel substrate. Spectrum was recorded in the positive ion mode. The resulting spectrum was the sum of 300 spectra obtained at different points of the sample. 2,5-Dihydroxybenzoic acid (DHB) (Acros, 99%) and α-cyano-4-hydroxycinnamic acid (HCCA) (Acros, 99%) were used as matrices.

**TGA.** Thermogravimetric analysis (TGA) was carried out in dynamic mode in 30 ÷ 600°C interval using a "Mettler Toledo TG50" system equipped with M3 microbalance allowing measuring the weight of samples in 0–150 mg range with 1 μg precision. Heating/cooling rate was chosen to be 10 °C/min. Every compound was studied twice: in the air and an under argon flow of 200 mL/min.

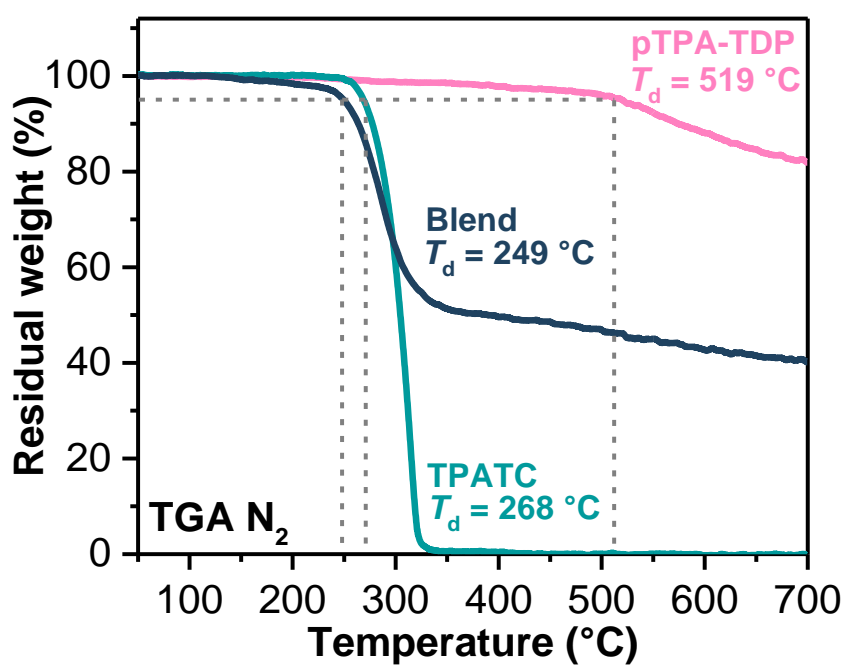

**Figure S3** – TGA curves in inert atmosphere of **pTPA-TDP**, **TPATC** and **Blend**

**DSC.** Differential scanning calorimetry (DCS) scans were obtained with a "Mettler Toledo DSC30" system with 20 °C/min heating/cooling rate in temperature range of +20–300 °C for all compounds. The $N_2$ flow of 50 mL/min was used.

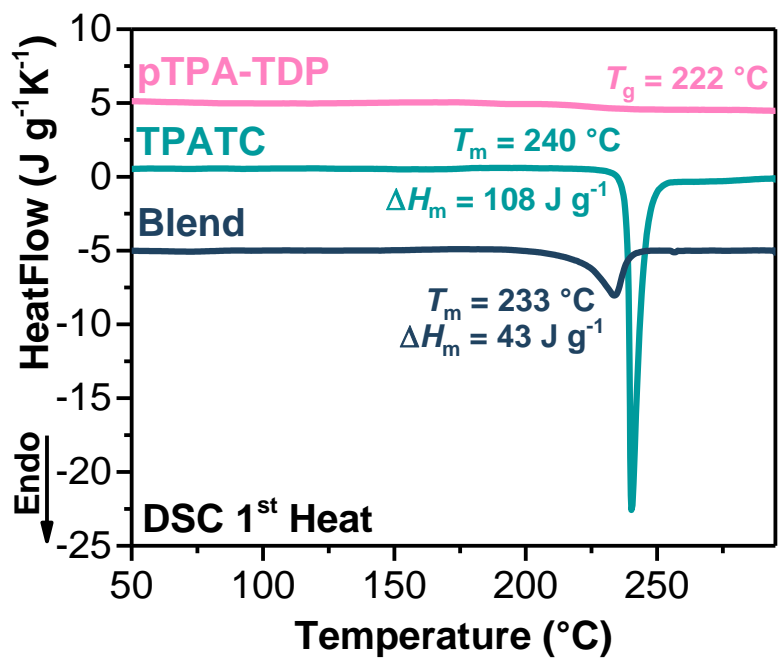

**Figure S4** – The first DSC heating scans of **pTPA-TDP**, **TPATC** and **Blend**

**UV-Vis steady state spectroscopy.** The absorption spectra of TPATC and pTPA-TDP and blended solutions were recorded with a "Shimadzu UV-2501PC" (Japan) spectrophotometer in the standard 10 mm photometric quartz cuvette using THF solutions of the corresponding compounds with the concentrations of $1 \times 10^{-5}$ mol L$^{-1}$. All measurements were carried out at RT. The optical properties in films were studied using a SE2030-010-DUVN spectrophotometer with a wavelength range of 200–1100 nm.

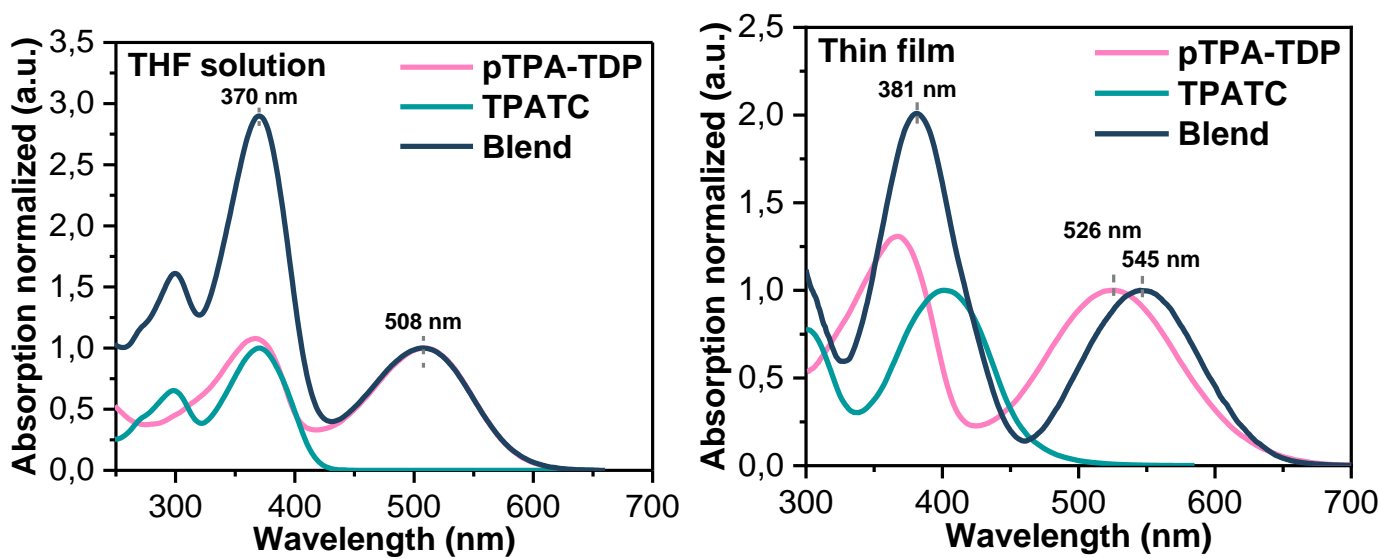

**Figure S5** - UV-Vis absorption spectra of **pTPA-TDP**, **TPATC** and **Blend** in THF solution and thin films cast from THF

**CV**. Cyclic voltammetry (CV) measurements for **TPATC** film were carried out with a three-electrode electrochemical cell in an inert atmosphere in an electrolyte solution, containing 0.1 M tetrabutylammonium hexafluorophosphate (Bu$_4$NPF$_6$) in an acetonitrile and 1,2-dichlorobenzene (4:1) mixture using IPC-Pro M potentiostat. The scan rate was 200 mV s$^{-1}$. The glassy carbon electrode was used as the work electrode. The film was applied to a glassy carbon surface used as a working electrode by rubbing. A platinum plate placed in the cell served as the auxiliary electrode. Potentials were measured relative to a saturated calomel electrode (SCE). The highest occupied molecular orbital (HOMO) and the lowest unoccupied molecular orbital (LUMO) energy levels were calculated using the first formal oxidation and reduction potentials, respectively, obtained from CV experiments in acetonitrile according to following equations: LUMO = –e($\varphi_{red}$+4.40) (eV) and HOMO = –e($\varphi_{ox}$+4.40) (eV) [S1,S2].

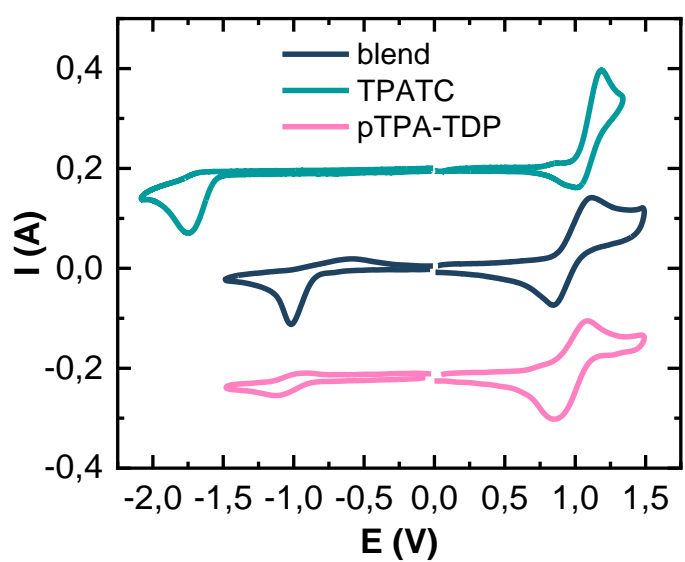

**Figure S6** - Cyclic voltammograms for thin polycrystalline film of TPATC, pTPA-TPD and blend in 1,2-dichlorobenzene/acetonitrile (1:4) mixture of solvents, which were recorded with a scan rate of 200 mV s$^{-1}$ using 0.1 M Bu$_4$NPF$_6$ as supporting electrolyte, glassy carbon (s = 2 mm$^2$) as work electrode, platinum plates as counter electrode and SCE (saturated calomel electrode) as reference electrode.

**Optical measurements.** The absorption spectra of **TPATC** and **pTPA-TDP** and blended solutions were recorded with a "Shimadzu UV-2501PC" (Japan) spectrophotometer in the standard 10 mm photometric quartz cuvette using THF solutions of the corresponding compounds with the concentrations of 1 × 10$^{-5}$ mol L$^{-1}$. All measurements were carried out at RT. The optical properties in films were studied using a SE2030-010-DUVN spectrophotometer with a wavelength range of 200–1100 nm. Time resolved photoluminescence measurements was performed with time correlated single photon counter technique (TCSPC) on Zolix OmniFluo-990 spectrofluorometer. The fluorescence was induced with 375 nm picosecond pulsed laser (CNILaser MDL-PS-375). The signal acquisition was conducted until 15000 counts.

**Film characterization.** The film thickness and uniformity were measured with Alpha-Step IQ profiler. The crystal structure of perovskite layers was investigated with X-ray diffractometer Tongda

TDM-10 using CuK$_α$ as a source with wavelength 1.5409 Å under 30 kV voltage and a current of 20 mA. Wetting angle was measured using KRÜSS EasyDrop DSA20. We used The SmartSPM 1000 system (NSG30 tip) in AC mode for scanning atomic force microscopy characterization of perovskite films. For KPFM measurements NSG10/Pt tips were used in NT-MDT Ntegra Prima. Tip was calibrated using fresh HOPG surface ($W_f^{HOPG}$ taken as 4.6 eV). Perovskite films absorbance spectra were studied using a SE2030-010-DUVN spectrophotometer with a wavelength range of 200–1100 nm.

*Device characterization.* JV-curves were measured in an ambient atmosphere by Keithley 2401 SMU with settling time of $10^{-2}$ s and voltage step of 24 and 85 mV for PSCs and PSMs respectively. The performance under 1 Sun illumination conditions were measured with ABET Sun3000 solar simulator (1.5 AM G spectrum, 100 mW/cm$^2$). Solar simulator was calibrated to standard conditions with a certified Si cell and an Ophir irradiance meter. The dark JV-curves measurements were performed in the dark box.

The EQE spectra were measured using QEX10 solar cell quantum efficiency measurement system (PV Measurements Inc., USA) equipped with xenon arc lamp source and dual grating monochromator. Measurements were performed in DC mode in the 300–850 nm range at 10 nm step. The system was calibrated using the reference NIST traceable Si photodiode. The conformity of spectral response for the measured PSCs was calibrated with Si-solar cell and was compliant to the ASTM E 1021-06 standard. The difference between Jsc values gathered from IV and those extracted from EQE measurements is mainly related to the different performances of the certified calibration cells used for adjustments to the standard conditions of illumination for the solar simulator and EQE system. The temperature measurements were done using lab-made thermostatic system equipped with 4 thermoelectric modules TEC1-12715 (12W), Arduino control and PID system with real-time temperature control (1°C precision). Maximum power point tracking (MPPT) was performed with the following algorithm (software was developed in LabVIEW): forward J–V scan; calculation of Pmax; Imax tracking at Vmax bias every 1 s and repeat of the cycle every 4 h. An LED projector calibrated to 100mWcm$^2$ with a certified Si photodiode to simulate 1 Sun conditions was used for illumination during MPPT. Ambient conditions were characterized using calibrated Luxmeter – LT-45 (480 -620 nm), insolation meter (500 – 2000 nm) Argus-03 (VNIIOFI), thermohydrometer RGK-30 from UNI-T.

*PIVTS & AS.*

In this work, we used paper employed a 470 nm, 250 mW/cm² LED with a 2 s light pulse duration to implement the Photoinduced Voltage Transient Spectroscopy (PIVTS) technique (Fig. X1). To verify the presence of mobile ions, we also performed Admittance Spectroscopy (AS). In this work, we used paper employed a 470 nm, 250 mW/cm² LED with a 2 s light pulse duration to implement the Photoinduced Voltage Transient Spectroscopy (PIVTS) technique. To verify the presence of mobile ions, we also performed Admittance Spectroscopy (AS). Notably, the high-power illumination was done through the optical window of the cryostat.

Such measurements are particularly sensitive for device active layer properties and its effect of mobile ions. High concentration of mobile charged defects can screen build-in of applied fields on Debye length scale leading to device structure capacitance changes: $(C = \frac{\epsilon\epsilon_0 A}{L_D}, L_D = \sqrt{\frac{\epsilon\epsilon_0 k_B T}{q^2 N_i}})$. The same approach used for determining times and frequencies for such defects will follow applied bias. For given diffusion coefficient $D$ time of mobile ion will travel across $L_D$ will be: $L_D^2/D = \tau$. So condition of peak in AS will be: $\omega \cdot \tau = 1 \Rightarrow \tau = \frac{\epsilon\epsilon_0 k_B T}{q^2 N_i D_0} \exp\left(\frac{E_A}{k_B T}\right)$, and then $E_A$ and $D_0$ can be determined from Arrhenius plot.

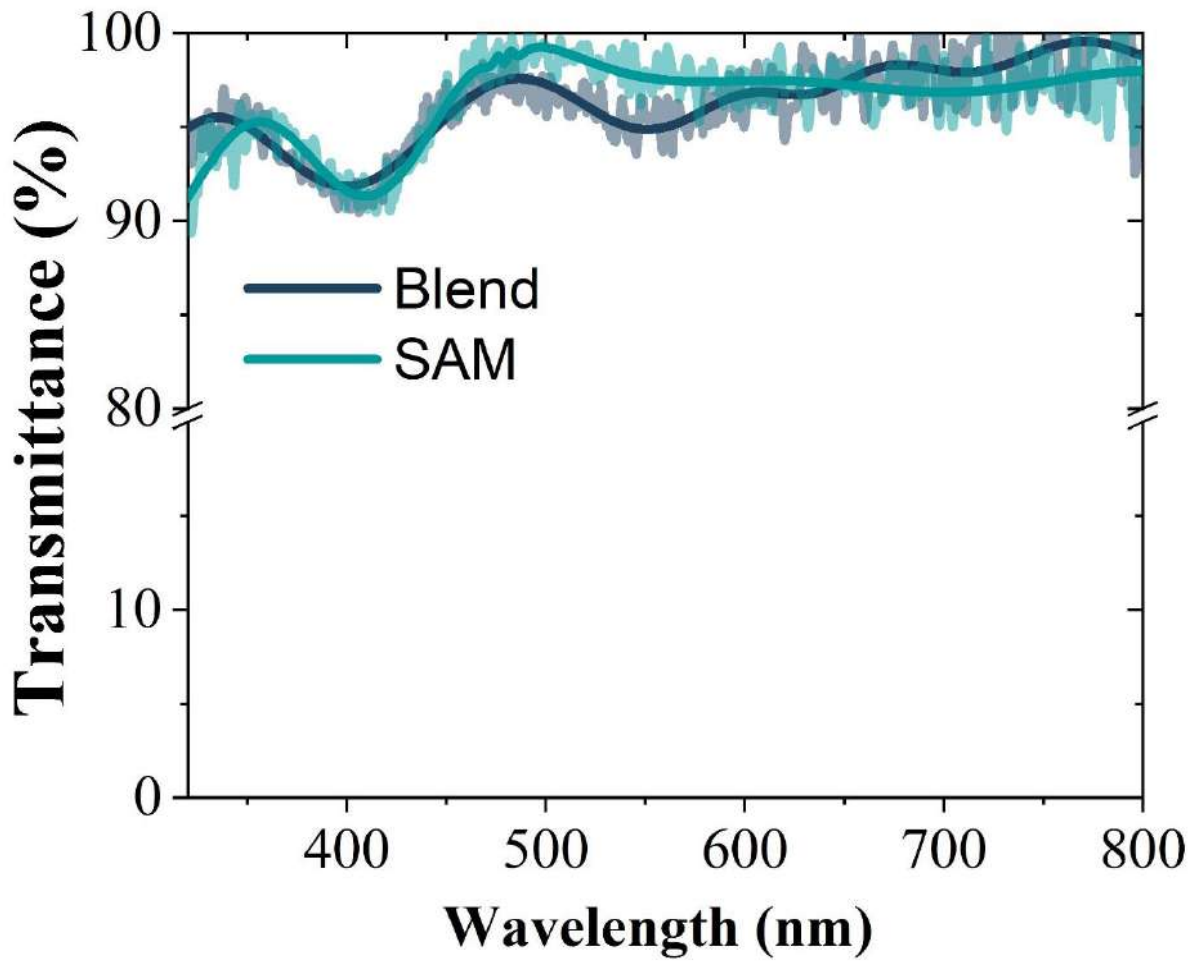

Figure S7 – Transmittance of the slot-die coated organic interlayers on the ITO-glass with NiO HTL

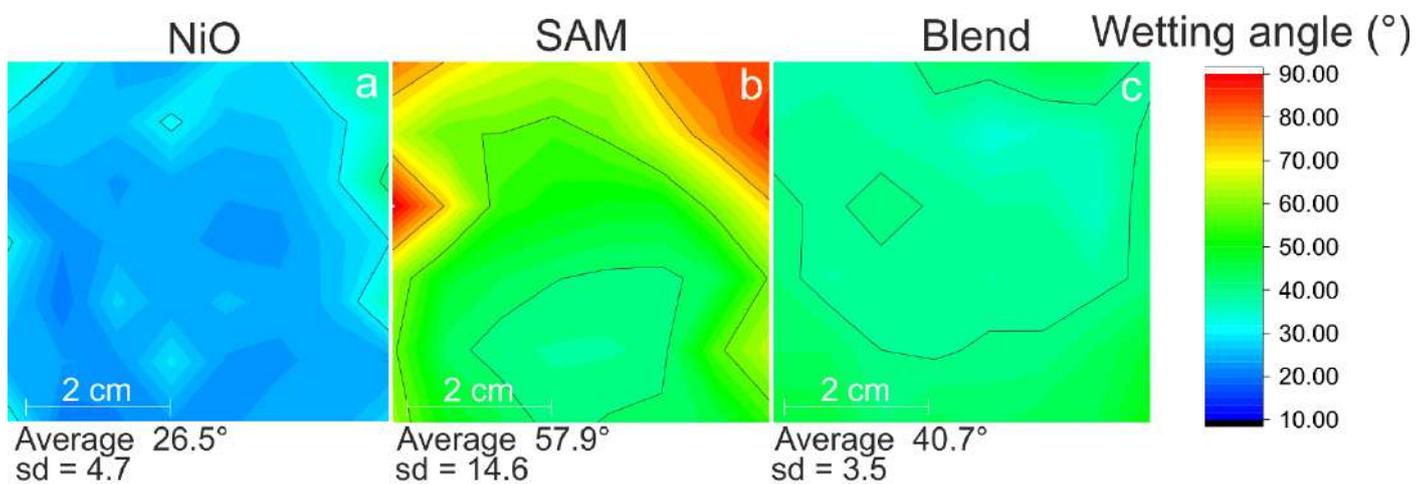

Figure S8 – The heat-map of CWA distribution across the 50 x 50 mm$^2$ ITO/NiO substrate (a) with pTPA-TDP (b) TPATC interlayer (c), and blended TPATC-pTPA-TDP (d)

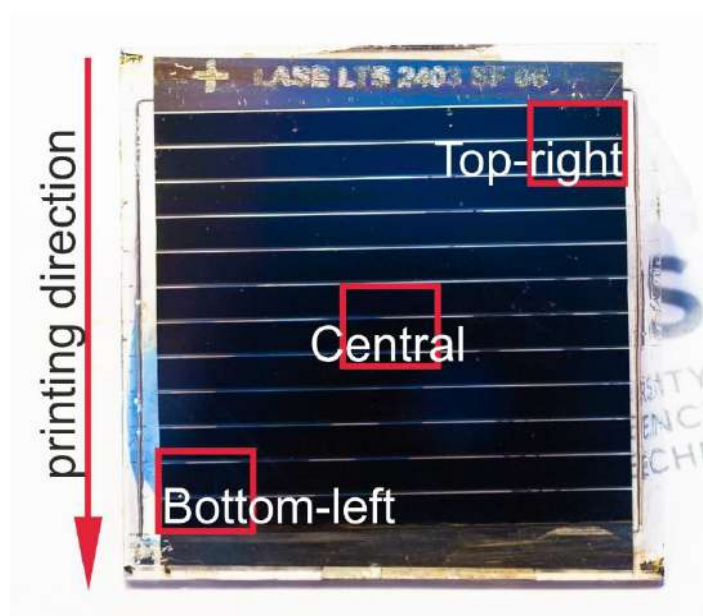

Figure S9 – choice of location on perovskite module for analysis

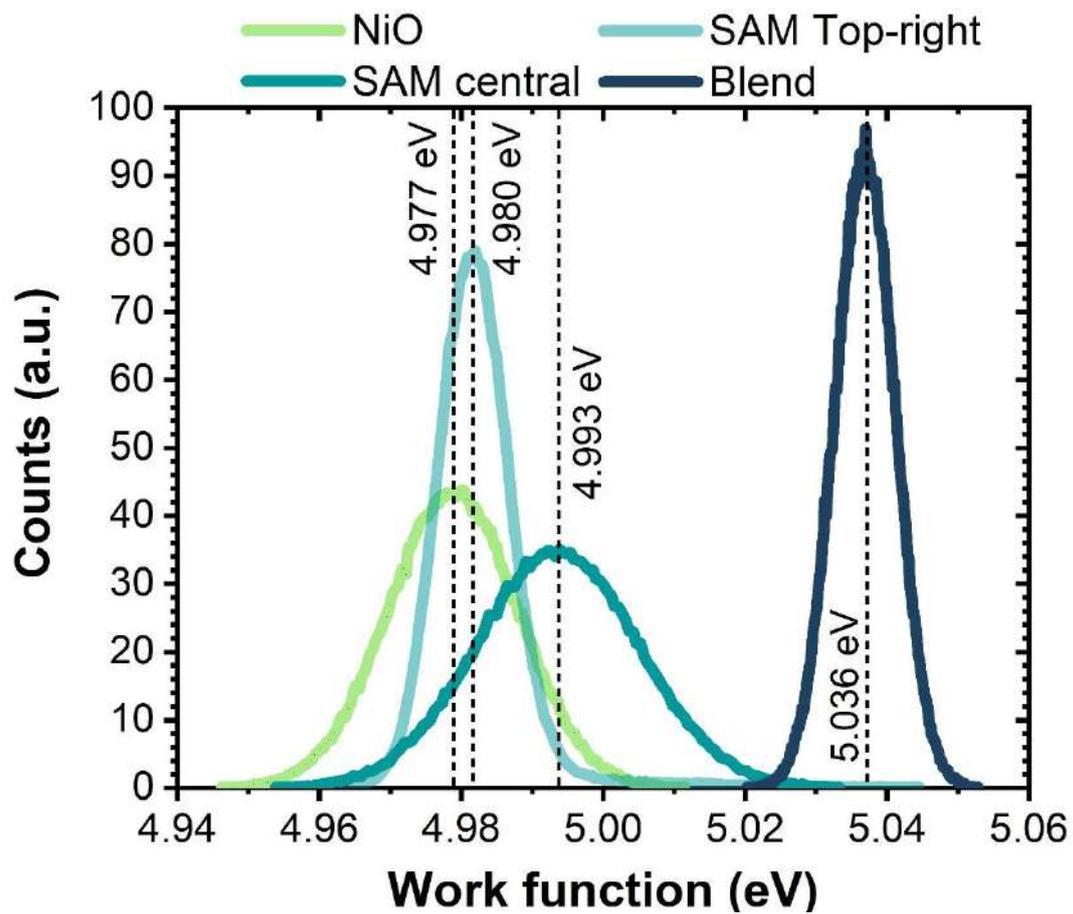

Figure S10 – $W_f$ distribution for NiO covered with organic passivation layer

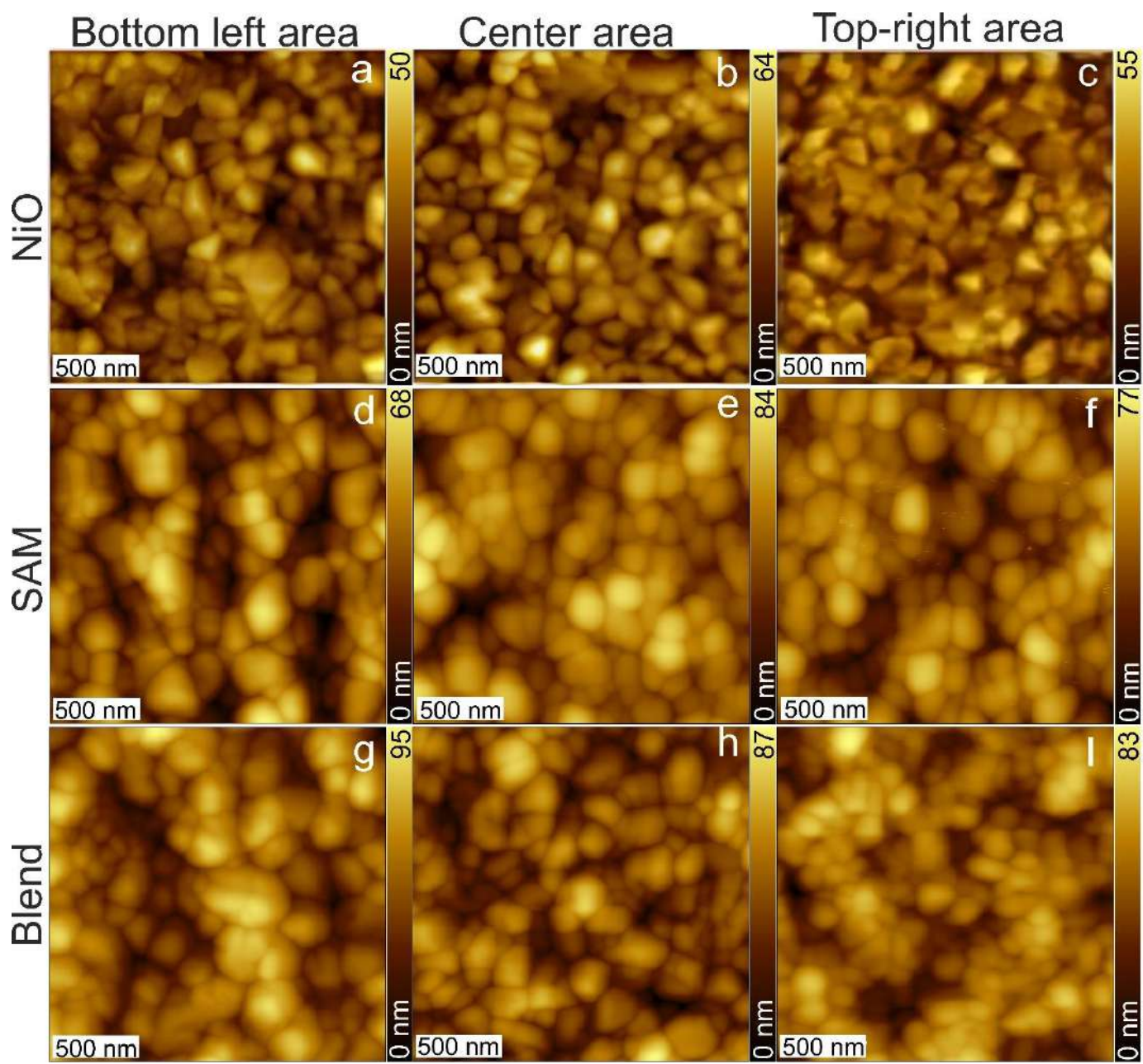

Figure S11 – AFM image of the slot-die coated perovskite absorber in the various HTL configurations

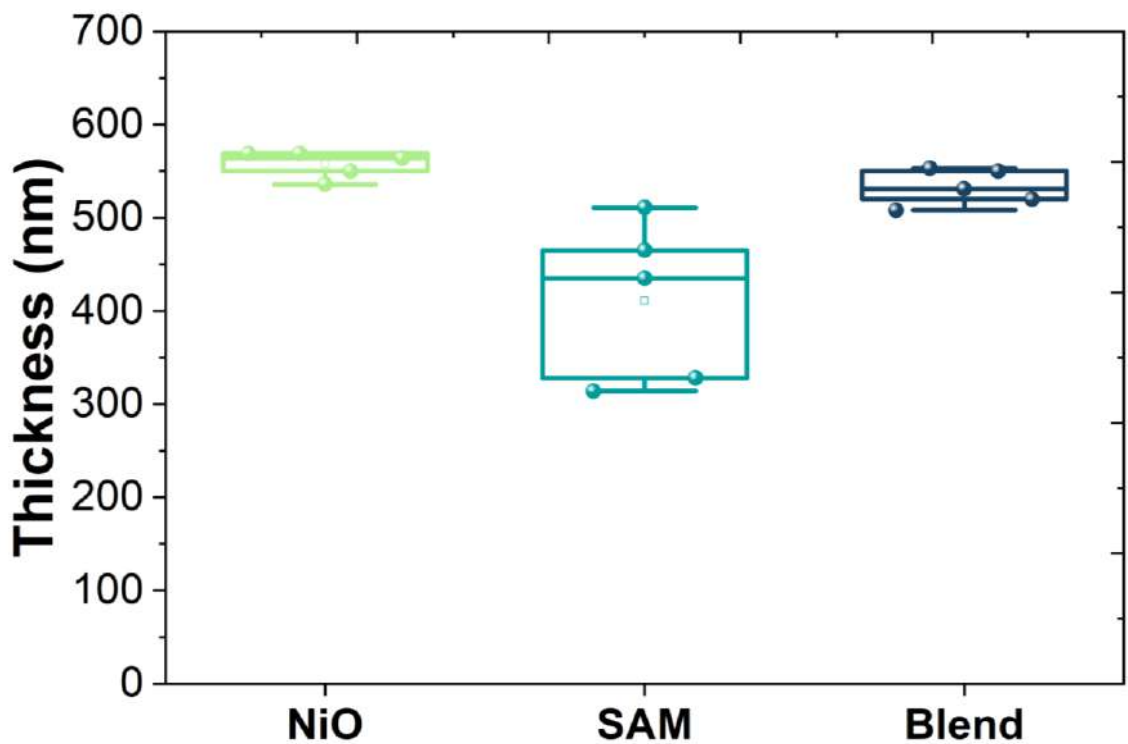

Figure S12 - The thickness values of perovskite absorber measured for various location on the 50x50 mm$^2$ substrate with different organic interlayers

**Table S1.** Thickness of slot-die coated TPATC and Blend films

| Organic interlayer | Average film thickness (standard deviation) |
|---|---|
| TPATC | 8.2 nm (1.1 nm) |
| Blend | 10.2 nm (1.5 nm) |

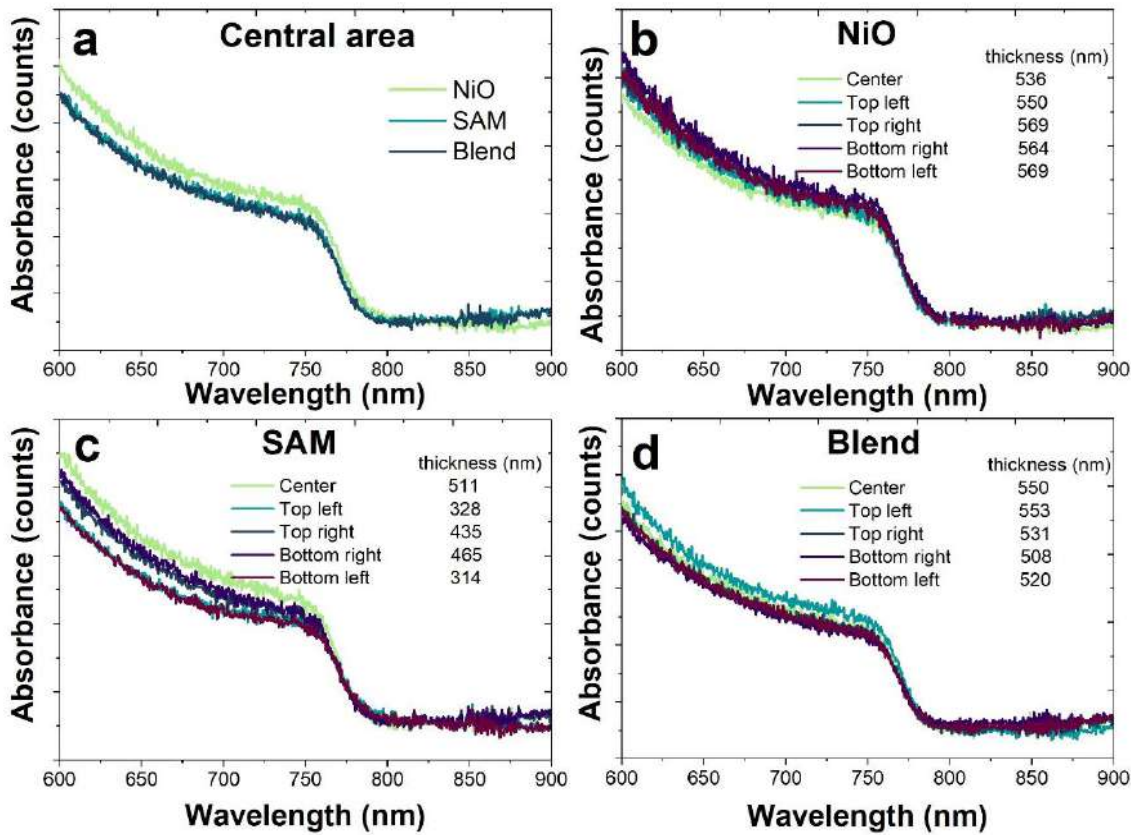

Figure S13 – absorbance spectra of slot-die coated perovskite films slot – die coated in the surface of the various HTL configurations

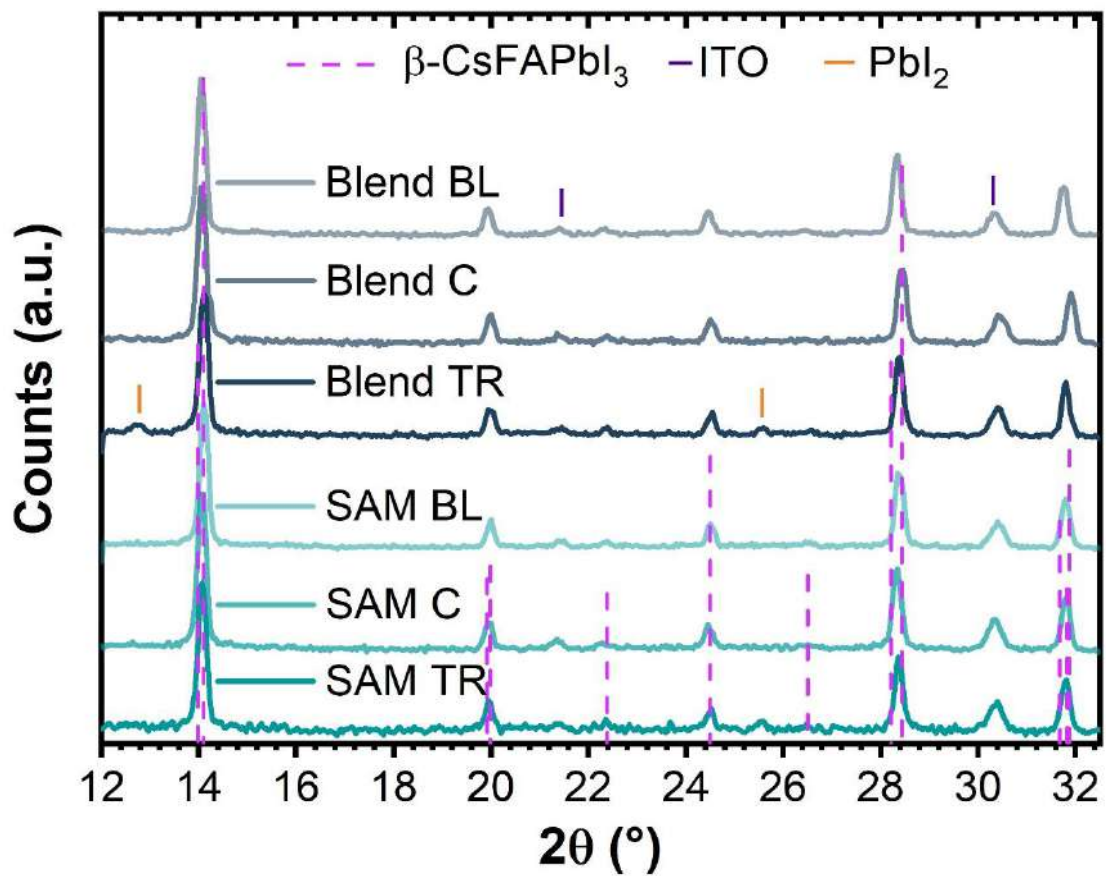

Figure S14 – perovskite film XRD deposited on passivation layer in Bottom-left (BL), Central (C), Top-right (TR) substrate locations

Table S2 – Lattice parameters of perovskite films in different substrate locations

| Sample | a, Å | c, Å |
|---|---|---|
| Reference (NiO) | 8.9217 | 6.3192 |
| SAM Bottom left | 8.9261 | 6.3052 |
| SAM Center | 8.9117 | 6.2777 |
| SAM Top right | 8.9250 | 6.3008 |
| Blend Bottom left | 8.9117 | 6.3007 |
| Blend center | 8.857 | 6.2659 |
| Blend top right | 8.8976 | 6.2943 |

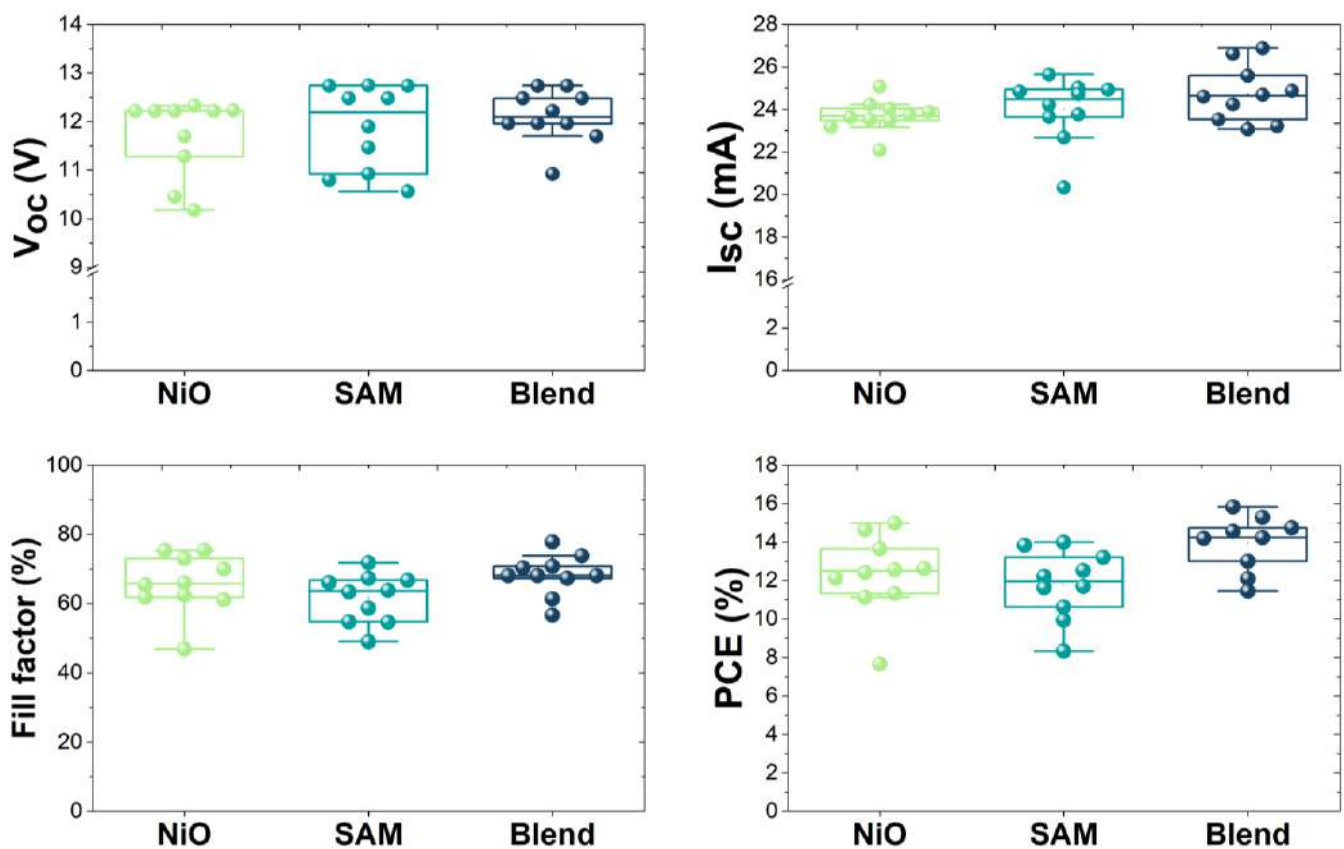

Figure S15 – Statistical distribution of the IV performance for the fabricated PSMs in different configurations

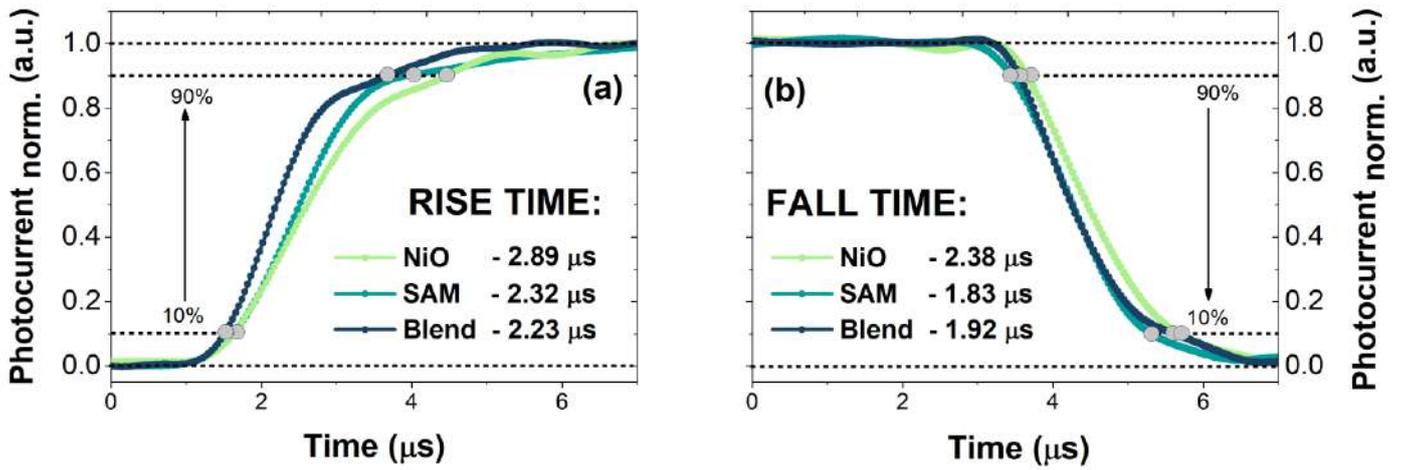

Figure S16 – Transient photo-current measurements for NiO, SAM and Blend configurations of slot-die coated PSCs in rise (a) and fall (b) modes

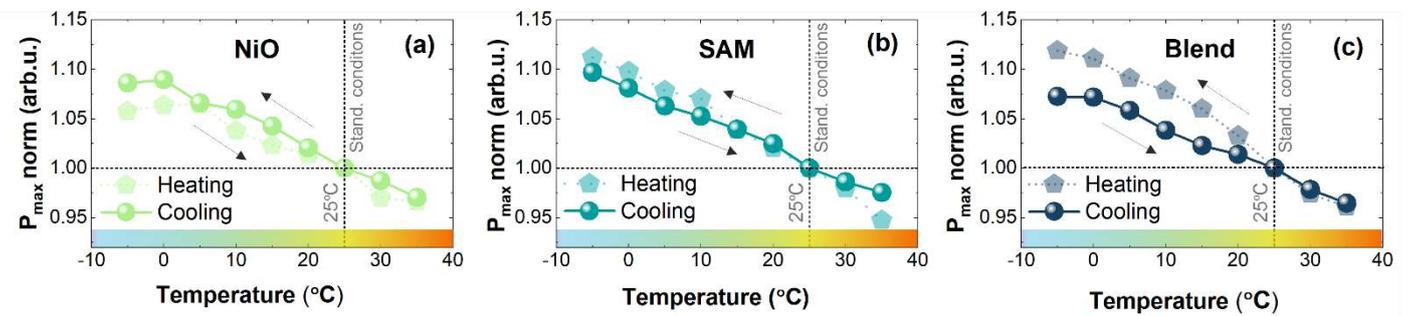

Figure S17 - The temperature dependence of the $P_{max}$ vs temperature for heating and cooling regimes

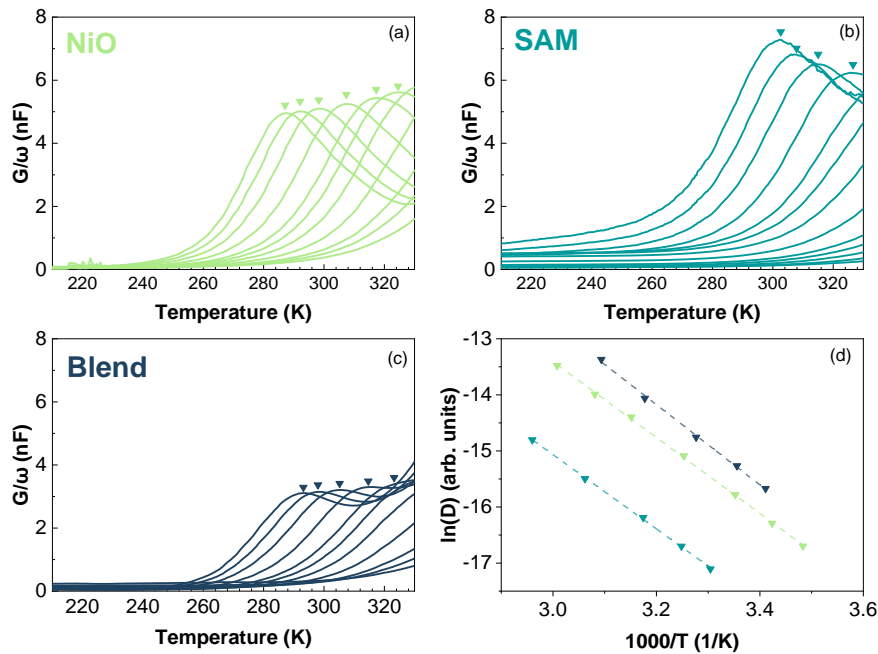

Figure S18 – (a)-(c) Results on Admittance Spectroscopy for studied samples with summurized data (d) on Arrhenius plot

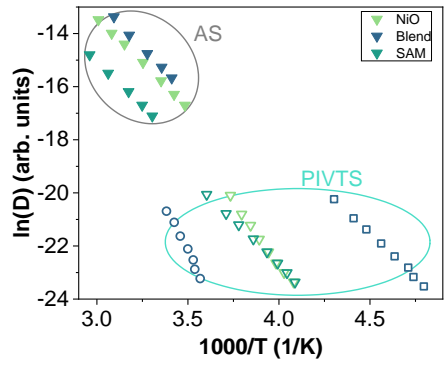

Figure 19 – All detected ions with AS and PIVTS techniques with labeling taken from [8,9],

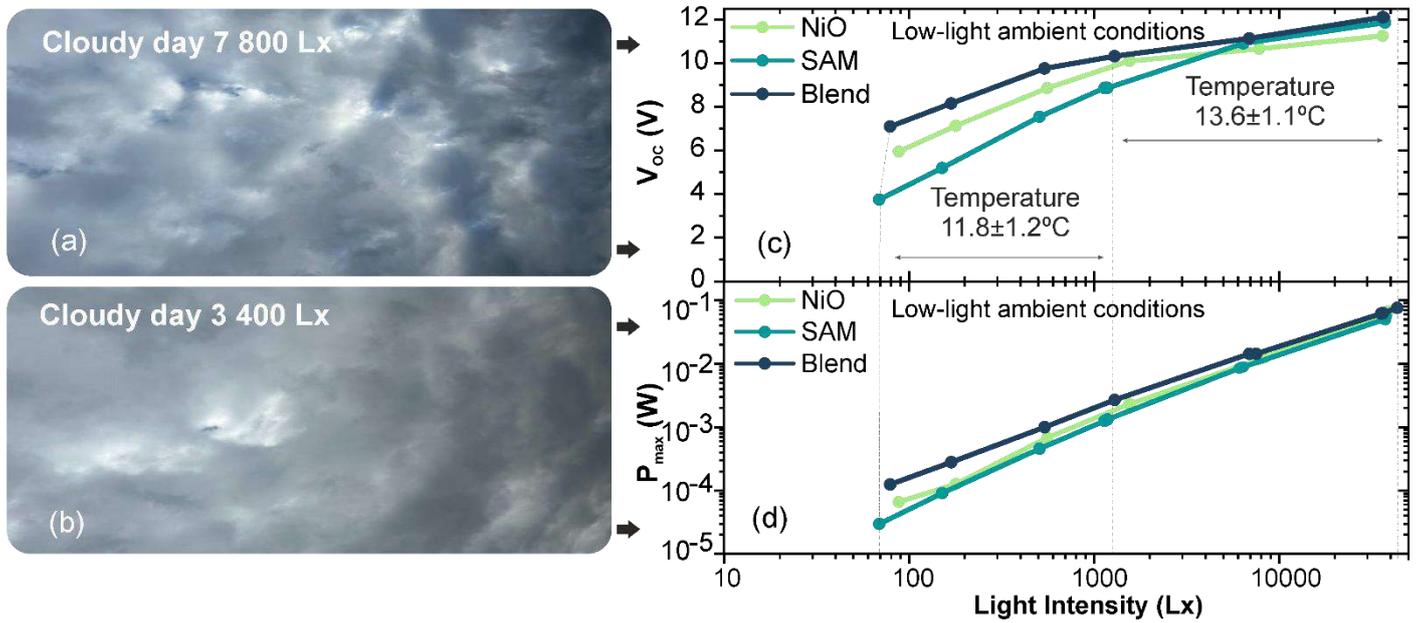

Figure S20 – The IV performance of PSMs with various slot-die coated organic interlayers measured under ambient conditions. The photo-image of the cloudy sky corresponding to the 7800 Lx intensity (a) The photo-image of the cloudy sky corresponding to the 3400 Lx intensity (b) The dependence of $V_{oc}$ measured in the ambient conditions (cloudy weather) vs light intensity (c) The dependence of $P_{max}$ (extracted from IV curves) in the ambient conditions (cloudy weather) vs light intensity (d)

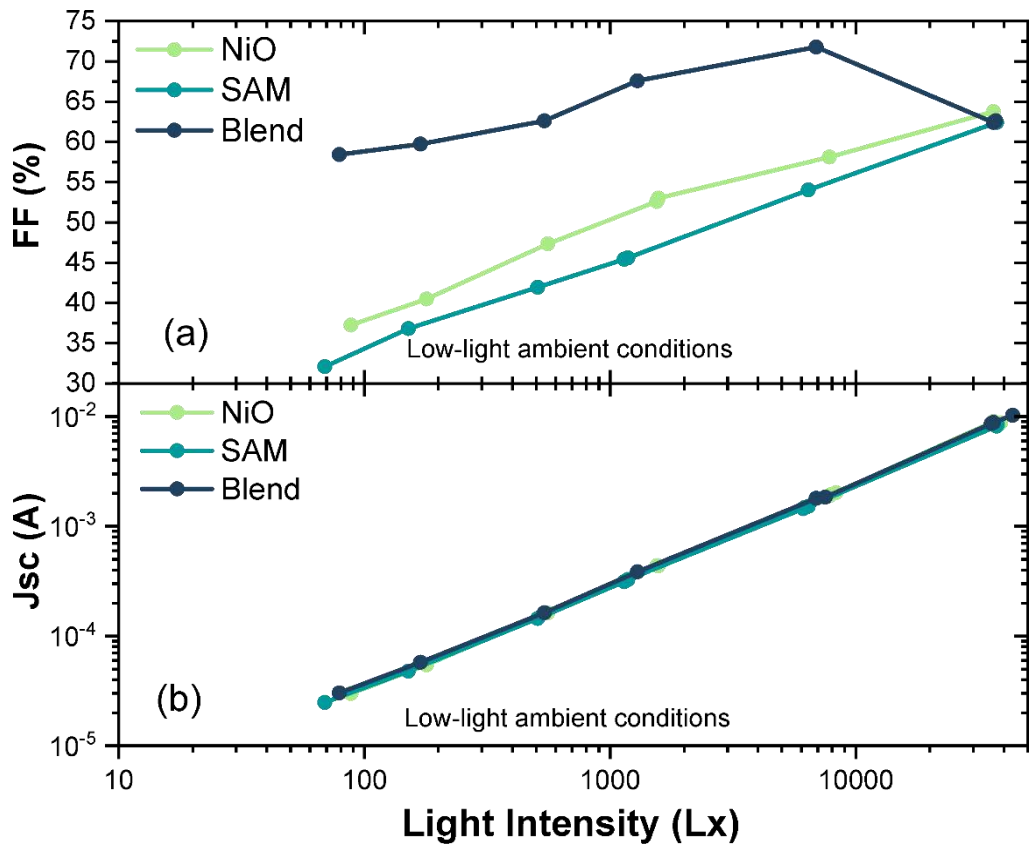

Figure S21 - The IV performance of PSMs with various slot-die coated organic interlayers measured under ambient conditions. The dependence of $J_{sc}$ measured in the ambient conditions (cloudy weather) vs light intensity (a), The dependence of FF measured in the ambient conditions (cloudy weather) vs light intensity (b)